\documentclass[11pt]{article}
\usepackage{aas_macros,amsmath,amssymb,comment,cite,esint,graphicx,mathtools,diagbox}
\usepackage{bm}
\usepackage[margin=.8in,letterpaper]{geometry}
\usepackage[colorlinks=true]{hyperref}
\usepackage[affil-it]{authblk}
\usepackage{subcaption}
\usepackage[utf8]{inputenc}
\usepackage{mathrsfs}
\usepackage{appendix}
\usepackage{amssymb}
\usepackage{float}                  
\usepackage{color}
\usepackage{cite}
\usepackage{hyperref}
\hypersetup{pageanchor=false}
\usepackage{indentfirst}
\usepackage{url}
\usepackage{xfrac}
\usepackage{caption}
\usepackage[numbers,square,comma,sort&compress,merge]{natbib}
\usepackage{esint}
\usepackage{overpic}
\usepackage{graphicx}
\usepackage{epsf,amsmath,bbold,amsfonts,stmaryrd}
\usepackage{textcomp}
\usepackage{ulem}
\usepackage{tikz}
\usepackage{multirow}
\numberwithin{equation}{section}
\usepackage{amsfonts}

\newcommand{\be}{\begin{equation}}
\newcommand{\ee}{\end{equation}}
\newcommand{\bea}{\setlength\arraycolsep{2pt} \begin{eqnarray}}
\newcommand{\eea}{\end{eqnarray}}
\newcommand{\nn}{\nonumber}
\newcommand{\mm}{\mathrm}
\newcommand{\mc}{\mathcal}

\def\ft#1#2{{\textstyle{\frac{\scriptstyle #1}{\scriptstyle #2} } }}
\def\fft#1#2{{\frac{#1}{#2}}}

\def\0{{\sst{(0)}}}
\def\1{{\sst{(1)}}}
\def\2{{\sst{(2)}}}
\def\3{{\sst{(3)}}}
\def\4{{\sst{(4)}}}
\def\5{{\sst{(5)}}}
\def\6{{\sst{(6)}}}
\def\7{{\sst{(7)}}}
\def\8{{\sst{(8)}}}
\def\sst#1{{\scriptscriptstyle #1}}

\begin{document}
\title{\bf Fast magnetic reconnection in Kerr spacetime}
\author{
Zhong-Ying Fan$^{1}$, Yuehang Li$^{2}$, Fan Zhou$^2$, Minyong Guo$^{2,3\ast}$}
\date{}

\maketitle

\vspace{-10mm}

\begin{center}
{\it
$^1$ Department of Astrophysics, School of Physics and Electronic Engineering, Guangzhou University, Guangzhou 510006, P. R. China\\\vspace{4mm}

$^2$ School of physics and astronomy, Beijing Normal University,
Beijing 100875, P. R. China\\\vspace{4mm}

$^3$Key Laboratory of Multiscale Spin Physics, Ministry of Education, Beijing 100875, P. R. China\\\vspace{4mm}


100871, P.R. China\\\vspace{4mm}

}
\end{center}

\vspace{4mm}

\begin{abstract}
We develop a relativistic scenario of fast magnetic reconnection process, for general magnetohydrodynamical plasmas around Kerr black holes. Generalizing the Petschek model, we study various properties of the reconnection layer in distinct configurations. When current sheet forms in the zero-angular-momentum (ZAMO) frame which corotates with the black hole, the reconnection rate for both radial and azimuthal configurations is decreased by spacetime curvature. However, when the current sheet forms in a non-ZAMO frame, which rotates either faster or slower than the black hole, detail analysis establishes that for any given slow rotations (subrelativistic at most) and mildly relativistic inflow, the ZAMO observer will find asymmetric reconnection rates for radial configuration: it is decreased on one side of the current sheet and is increased on the other side in comparison to the unrotation limit. This is valid to both the Sweet-Parker and the Petschek scenario. The results clarify the effects of rotation on the reconnection layer in the laboratory frame in the flat spacetime limit. 
\end{abstract}

\vfill{\footnotesize $\ast$ Corresponding author: minyongguo@bnu.edu.cn;}

\thispagestyle{empty}

\pagebreak

\tableofcontents
\addtocontents{toc}{\protect\setcounter{tocdepth}{2}}




\section{Introduction}

Magnetic reconnection is a fundamental process for magnetohydrodynamical plasmas, which converts magnetic energy into plasmas particle energy rapidly. It is generally considered to be a competitive mechanism to explain many high energy phenomenons in astrophysics \cite{book,biskamp}, such as coronal mass ejections, magnetospheric substorms, stellar flares and gamma-ray bursts. 

Despite that magnetic reconnection was usually studied in the non-relativistic regime \cite{book,biskamp}, it was recently recognized that special relativistic effect is of paramount importance for strongly magnetized plasmas \cite{Zenitani:2001fef,Lyubarsky:2008yi}. This is the case when the magnetic energy density exceeds the plasmas rest mass density greatly so that the Alfv\'{e}n wave speed approaches to the speed of light. Generalization of magnetic reconnection to relativistic regimes has been conducted by a number of authors, both analytically \cite{Blackman1994,Lyutikov:2002bt,Lyubarsky:2005zt,Comisso:2014nva} and numerically 
\cite{Zenitani:2009bj,Liu:2014ada,Zenitani:2010vq,Takahashi:2011br}. It was found that relativistic effect is extremely important for magnetic energy conversion and particle accelerations \cite{Kowal:2012rv,Kowal:2011yz,Zenitani:2007vt,Cerutti:2011cv}. In particular, the highly efficient conversion of magnetic energy into nonthermal particle energy
\cite{Guo:2014via,Sironi:2014jfa,Guo:2015cua, Chen:2024ggq} suggests that relativistic reconnection is a primary candidates to explain the high energy emissions in magnetically dominated environments like pulsars and jets from gammar-ray bursts and from active galactic nuclei. 

In contrast, the effects of spacetime curvature of black holes to magnetic reconnection are much less explored. In \cite{Asenjo:2017gsv}, generalization of Sweet-Parker model to curved spacetime was first developed by Asenjo and Comisso. It was established that for various current sheet configurations, the reconnection rate will be decreased by the spacetime curvature in comparison to the flat spacetime limit. Inclusion of collisionless effect and gravitational electromotive force for magnetic reconnection in curved spacetime are further explored in \cite{Comisso:2018ark,Asenjo:2019nji}. However, for the Sweet-Parker scenario, an essential equetion is the reconnection rate is very small (without collisionless effect): it is of the order $\sim S^{-1/2}$, where $S=L/\eta$ is the relativistic Lindquist number. Since the Lindquist number is generally very large in astrophysical environments, the characteristic time scale of energesis in this model is usually much larger than the observation time scale. It is widely known that this issue can be resolved in a Petschek like model, in which the reconnection rate is significantly improved to $\sim (\mm{ln}{S})^{-1}$. 

In this work, our first motivation is to develop the Petschek model for relativistic magnetohydrodynamic plasmas around rotating black holes. We will study the effects of spacetime curvature to the reconnection rate and other important properties of the reconnection layer for various current sheet configurations. Our second motivation is to study magnetic reconnection for a general rotating reconnection layer. It was assumed in \cite{Asenjo:2017gsv} that plasmas corotates with the black hole and the reconnection was studied for the zero-angular-momentum (ZAMO) observers. Since generically the current sheet may form in a non-ZAMO frame which rotates either faster or slower than the ZAMO frame, we would like to study the general situation from the view of the ZAMO observer carefully. We explore a general rotating disk in the equatorial plane of the black hole. We will show that for slow rotations (subrelativistic at most) and mildly relativistic inflow, the ZAMO observer will always find asymmetric reconnection rates for radial configuration: it is decreased on one side of the current sheet and is increased on the other side in comparison to the unrotation limit.

The paper is organized as follows. In section \ref{sec2}, we introduce the governing equations for relativistic magnetohydrodynamical plasmas in curved spacetime. In section \ref{sec3}, we explore various properties of Petschek model for radial current sheet configuration in the ZAMO frame. We compute the reconnection rate and establish that the gravitational potential gives the leading order corrections to decrease the reconnection rate in the weak gravity regimes. In section \ref{sec4}, we study various properties of the reconnection layer in azimuthal direction for ZAMO observers. In section \ref{sec5}, we study both the Sweet-Parker and the Petschek model in the non-ZAMO frame. We focus on a rotating reconnection layer in radial configuration. We briefly conclude in section \ref{sec6}.

\section{Preliminaries}\label{sec2}
Consider relativistic magnetohydrodynamic (MHD) fluids in curved spacetime.  The fluids is governed by the continuity equation \cite{Lichnerowicz}
\be \nabla_\mu(\rho U^\mu)=0 \,,\ee
the energy-momentum conservation
\be \nabla_\nu(h U^\mu U^\nu)=-\nabla^\mu p+J^\nu F^\mu_{\,\,\,\nu} \,,\label{emcon}\ee
and the Ohm's law
\be U^\nu F^\mu_{\,\,\,\nu}=\eta(J^\mu-\rho_e U^\mu) \,,\ee
where $\rho (\rho_e)$ is the mass (charge) density, $h$ is the enthalpy density, $p$ is the pressure, $U^\mu$ is the fluid four velocity and $J^\mu$ is the electric current. Besides, MHD fluids obeys the Maxwell's equation
\be \nabla_\nu F^{\mu\nu}=J^\mu \,.\ee
These provide a complete set of equations for MHD in a fixed background. 

Consider rotating black holes, whose metric can be written in the form of
\be ds^2=-\alpha^2 dt^2+\sum_{i=1}^3\big( h_idx^i-\alpha \beta^i \big)^2 \,,\ee
where $h_0^2=-g_{00}\,,h_i^2=g_{ii}$. Here $\beta^i$ is related to the black hole rotations $\beta^i=h_i\omega_i/\alpha\,,\omega_i=-g_{0i}/h_i^2$ and $\alpha^2=h_0^2+\sum_i (h_i\omega_i)^2$. An advantage of this coordinate is one can introduce the ZAMO frame $\{\hat t\,,\hat x^i\}$ \cite{Bardeen:1972fi}, under which the spacetime is locally flat $ds^2=\eta_{\mu\nu}d\hat{x^\mu}d{\hat{x^\nu}}$. One has
\be d\hat{t}=\alpha dt\,,\qquad d\hat{x}^i=h_idx^i-\alpha\beta^i dt \,,\ee
or the inverse transformation
\be dt=\alpha^{-1}d\hat{t}\,,\qquad dx^i=h_i^{-1}\big(d\hat{x}^i+\beta^i d\hat t\big) \,.\ee
Throughout out this paper, we use hats to stand for all vectors or tensors in the ZAMO frame.

To study magnetic reconnection in the ZAMO frame, we rewrite the energy-momentum equation (\ref{emcon}) in the form of 
\be \nabla_\nu S^{\mu\nu}=H^\mu \,,\ee
where
\bea
&& S^{\mu\nu}=h U^\mu U^\nu\,,\nn\\
&& H^\mu=-\nabla^\mu p+J^\nu F^\mu_{\,\,\,\nu}\,.
\eea
It was established in \cite{Koide:2009dt} that any equation of this form in the ZAMO frame can be expressed into
\bea
&& \fft{1}{h_1h_2h_3}\sum_{j=1}^3\fft{\partial}{\partial x^j}\Big[ \fft{\alpha h_1h_2h_3}{h_j}\big(\hat{S}^{ij}+\beta^j \hat{S}^{i0} \big) \Big]+\fft{\hat{S}^{00}}{h_i}\fft{\partial\alpha}{\partial x^i} \nn\\
&&-\alpha\sum_{j=1}^3\big( G_{ij}\hat{S}^{ij}-G_{ji}\hat{S}^{jj}+\beta^jG_{ij}\hat{S}^{0i}-\beta^jG_{ji}\hat{S}^{0j} \big)\nn\\
&&+\sum_{j=1}^3\sigma_{ji}\hat{S}^{0i}=\alpha \hat{H}^i \,,
\eea
where $G_{ij}=-(1/h_ih_j)\partial h_i/\partial x^j$ and $\sigma_{ij}=-(1/h_j)\partial(\alpha\beta^i)/\partial x^j$. 

Since we are interested in studying magnetic reconnection in Kerr black holes, we present the metric in the Boyer-Lindquist (BL) coordinates $(x^0\,,x^1\,,x^2\,,x^3)=(t\,,r\,,\theta\,,\phi)$
\bea
&& h_0=\sqrt{1-\fft{2r_g r}{\rho^2}}\,,\qquad h_1=\fft{\Sigma}{\sqrt{\Delta}}\,,\nn\\
&&h_2=\Sigma\,,\qquad h_3=\fft{\Pi\sin\theta}{\rho}\,,
\eea
where $r_g=GM$ is the gravitational radius and $a=J/J_{\mm{max}}\leq 1$ is the rotation parameter ($J$ is the angular momentum and $J_{\mm{max}}=GM^2$). The functions $\rho\,,\Delta$ and $\Pi$ are given by
\bea
&&\Sigma^2=r^2+a^2r_g^2\cos^2\theta\,,\nn\\
&&\Delta=r^2-2r_gr+a^2r_g^2\,,\nn\\
&&\Pi^2=\big(r^2+a^2r_g^2 \big)^2-\Delta\big(ar_g\sin\theta \big)^2\,.
\eea
In addition, the black hole rotates only around the $\phi$-direction 
$\omega_1=\omega_2=0\,,$ and $\omega_3=2r_g^2 ar/\Pi^2$ so that $\beta^i=\beta^\phi \delta^{i\phi}$.

To study magnetic reconnection in a Petschek-like scenario, we consider a quasi-two-dimensional electric current sheet (with the thickness far less than the length $\delta \ll L$) and adopt quasi-stationary conditions ($\partial_t\approx 0$). While the current sheet can form in different locations around the black holes, we will focus on two configurations in the equatorial plane of the black hole.  We will assume the current sheet is comoving with the ZAMO frame at first. The situation of a general rotating reconnection layer will be studied later.

\section{Reconnection layer in radial direction}\label{sec3}

\begin{figure}
\centering
\includegraphics[width=350pt]{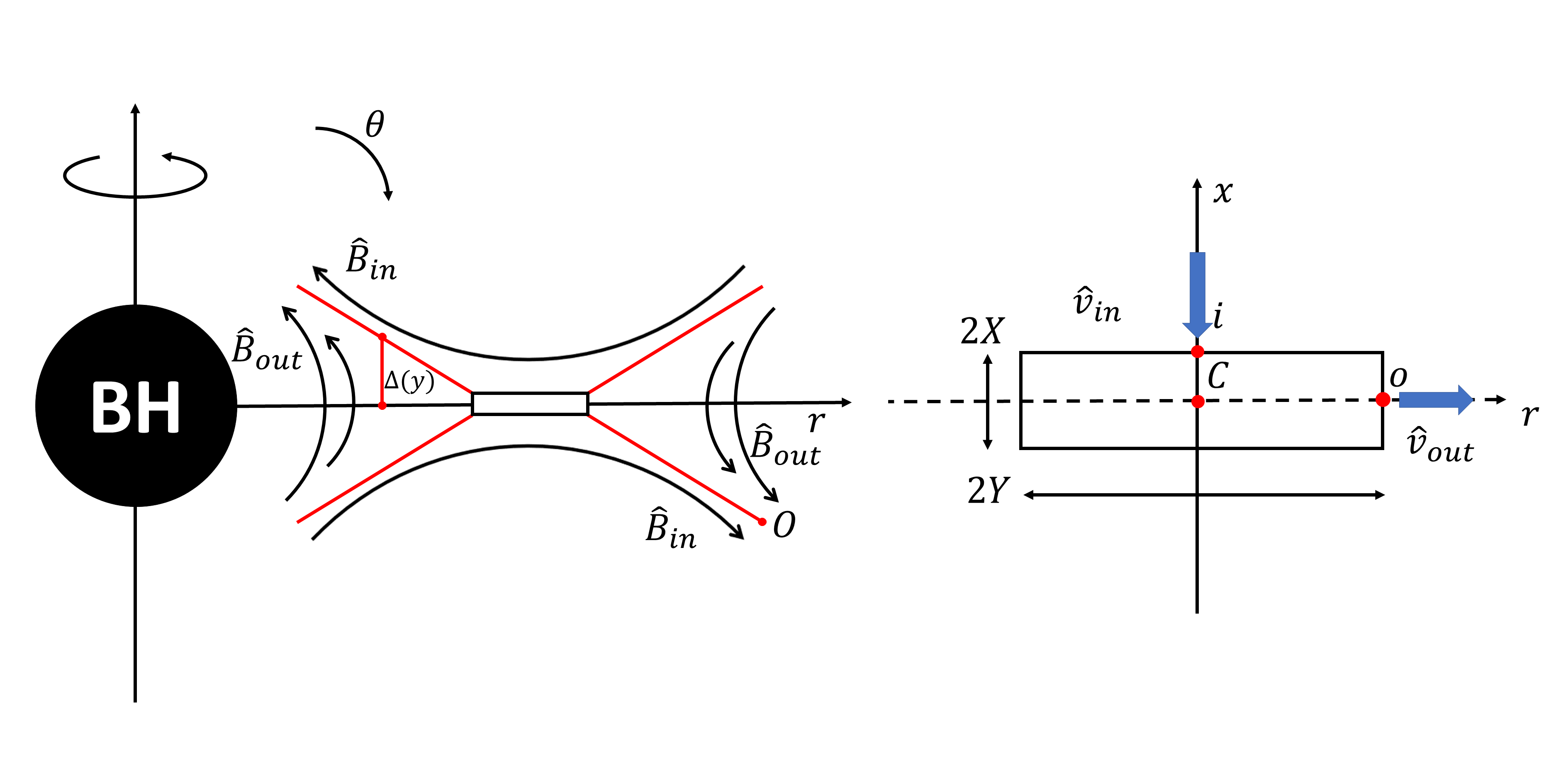}
\caption{Raidal magnetic reconnection layer of Petschek model in Kerr spacetime. Here $y\equiv r-r_C$.}
\label{radialpetschek}
\end{figure}

Consider reconnection layer in the radial direction at first, see the illustration figure Fig. \ref{radialpetschek}. In a Petschek model, there are three patches in the current sheet: i) The inflow region. The plasma is viewed as the ideal MHD fluid. In particular, the ingoing magnetic fields upstream of the sheet are not anti-parallel any longer. The field lines enter the diffusion region with an inclination angle. This is extremely important. The magnetic field line bends down and is reconnected at the neutral line. It turns out that even if the magnetic field has small perturbations around a uniform background, the reconnection rate will be improved significantly. ii) The diffusion region. In this region, magnetic reconnection develops via Sweet-Parker process. Previously this region was studied carefully in \cite{Asenjo:2017gsv}. The reconnection rate is evaluated as
\be \mc{M}=\hat{v}_i\sim \fft{X}{Yh_1}\Big|_{C}\sim \Big(\fft{Y h_1}{\eta}\Big)^{-1/2}\Big|_C \,.\label{diffusionrate}\ee
However, in a Petschek model, this region is extremely small, compared to the sheet length $Y\ll L$.
iii) The outflow region. In this region, the shape of the current sheet is generally complicated, with the thickness depending on the radial coordinates. Consider an arbitrary point $O$ on the boundary $\Delta=\Delta(y)\,,y=r-r_C$. The boundary approaches asymptotically to a straight line, as will be shown later.

To proceed, we work in the quasi-stationary limit $\partial_t\approx 0$. Since the sheet is put in the radial direction, we assume $\hat{v}^\phi\approx 0 \approx \hat{B}^\phi$ and $\partial_\phi\approx 0$. The electric current and the reconnected electric field are given by $J=\hat{J}^\phi e_\phi\,,E=\hat{E}^\phi e_\phi$, where $e_\phi$ stands for unit vector in azimuthal direction\footnote{In the language of differential geometry, the unit vector is represented by $e_x=\ft{\partial}{\partial x}$}. Without loss of generality, we focus on the upper half region of the sheet. We set $\hat{v}_{\mm{in}}=-\hat{v}_i e_\theta$ and $\hat{B}_{\mm{in}}=-\hat{B}_0 e_{r}+\hat{B}_1$, where $B_0$ is a constant magnetic field. Since we are interested in small perturbations around a uniform magnetic field, we assume $|\hat{B}_1|\ll \hat{B}_0$. The outflow velocity $\hat{v}_o$ in the ZAMO frame can be calculated using the $r$-component of the energy-momentum equation
\be \fft{\partial}{\partial r}\Big(h\hat{\gamma}^2\big(\hat{v}^r \big)^2 \Big)+\fft{\partial}{\partial \theta}\Big(h_1h_2^{-1}h\hat{\gamma}^2 \hat{v}^r \hat{v}^\theta \Big)+h\hat{\gamma}^2\fft{\partial\mm{ln}\alpha}{\partial r}=-\fft{\partial p}{\partial r}-h_1\hat{J}^\phi\hat{B}^
\theta \,.\label{emeqr}\ee
Notice that the second term on the l.h.s cannot be dropped since $O$ is located on the boundary. This term gives nontrivial contributions when integrating the equation along $\theta$-direction. Notice that we consider a short current sheet $L\ll r_g$ and focus on weakly gravity regimes so that gravitational tidal force could be ignored when performing integrations (however when collisionless effect is included, gravitational electromotive force will be important and cannot be ignored even in the weak gravity regimes \cite{Asenjo:2019nji}). In principle, integrating the equation from $C$ to $O$, we are able to derive the outflow velocity but the result depends on the sheet thickness $\Delta(y)$ and the magnetic field $\hat{B}^{\theta}$ in the outflow region. So let's first study general behaviors of these quantities.

The first estimation of the thickness $\Delta(y)$ can be achieved from flux conservation. The inflow flux $\partial_\theta \big(\alpha h_1h_3 \rho\hat{\gamma}_i \hat{v}_i \big)/(\alpha h_1h_2h_3)\approx\rho\hat{\gamma}_i \hat{v}_i/\Delta(y)$ must balance the outflow flux\\
$\partial_r \big(\alpha h_2h_3 \rho\hat{\gamma}_o \hat{v}_o \big)/(\alpha h_1h_2h_3)\approx\rho\hat{\gamma}_o \hat{v}_o/\big(y h_1(y)\big)$. This gives
\be \Delta(y)\approx \fft{\hat{\gamma}_i\hat{v}_i}{\hat{\gamma}_o\hat{v}_o}\,yh_1(y) \,.\label{delta1}\ee
Next, the Ohm's law on the current sheet leads to
\be \hat{\gamma} \hat{E}^\phi+\hat{\gamma}\hat{v}^r\hat{B}^\theta=\eta \hat{J}^\phi \,.\label{ohmradial}\ee
Since beyond the sheet the plasmas is ideal MHD fluid $\eta\approx 0$, the electric field at the center incident point is $\hat{E}^\phi|_i\approx \hat{v}_i \hat{B}_0$.  
Quasi-stationary approximation implies the electric field on the sheet is homogeneous so $\hat{E}^\phi_o\approx \hat{E}^\phi|_i$. In addition, the electric current $\hat{J}^\phi$ at $O$ can be evaluated from the Maxwell's law
\be \hat{J}^\phi\Big|_o\approx h_2^{-1}\partial_\theta \hat{F}^{\phi\theta}\approx \fft{\hat{B}_0}{\Delta(y)} \,.\ee
Integrating (\ref{ohmradial}) along thickness of the sheet and using (\ref{delta1}), we deduce 
\be \Delta(y)-\fft{\hat{\gamma}_i\hat{B}_o}{\hat{\gamma}_o \hat{B}_0}\,y h_1(y)-\fft{\eta}{\hat{\gamma}_o \hat{v}_i}=0 \,,\label{master1}\ee
where we have set $\hat{B}^\theta\Big|_O\equiv -\hat{B}_o$. This connects thickness of the sheet to the reconnected magnetic field on the boundary. Close to the diffusion region, the second term can be dropped so that $\Delta\approx \eta/\hat{v}_i\approx X$, consistent with the Sweet-Parker process \cite{Asenjo:2017gsv}. Otherwise, in the far outflow region $y\approx L$, the third term can be dropped so that asymptotically
\be \fft{\Delta(y)}{y h_1(y)}\Big|_{y\approx L}\approx \fft{\hat{\gamma}_i \hat{B}_o}{\hat{\gamma}_o \hat{B}_0} \,.\label{deltafar}\ee
Combing the result with (\ref{delta1}), one finds the reconnection rate is related to the asymptotic thickness or the magnetic field
\be \hat{v}_i\approx \fft{\hat{B}_o}{ \hat{B}_0} \approx \fft{\Delta(y)}{y h_1(y)}\Big|_{y\approx L}\,,\label{rate1}\ee
where in the second equality we have assumed the outflow is mildly relativistic $\hat{\gamma}_o\approx 1$, which will be verified later. This tells us the boundary approaches asymptotically to a straight line, with a rate $\hat{v}_i$ . 

To determine the full shape of the boundary, we integrate the momentum equation (\ref{emeqr}) along the sheet thickness and obtain
\be \fft{\hat{B}_o}{\hat{B}_0}=\fft{h\hat{v}_i^2}{\hat{B}_0^2}\Big[ 2y\partial_y\big( \fft{y h_1}{\Delta} \big)+\hat{\gamma}_o \fft{yh_1}{\Delta} \Big]+\fft{h\hat{\gamma}_o^2 }{\hat{B}_0^2 }\,\fft{\Delta}{yh_1}\,y\partial_y\mm{ln}\alpha \,,\label{master2}\ee
where we have adopted (\ref{delta1}). For hot relativistic plasmas $h=4p$ and at the neutral line, pressure balance implies $p=\hat{B}_0^2/2$. 
By plugging this equation into (\ref{master1}), we deduce
\be \big(1-2\hat{\gamma}_o \,y\partial_y\mm{ln}\alpha \big)\Delta(y)-2\hat{v}_i^2yh_1\Big[2\hat{\gamma}_o^{-1}y\partial_y\big( \fft{y h_1}{\Delta} \big)+\fft{yh_1}{\Delta}\Big]-\fft{\eta}{\hat{\gamma}_o\hat{v}_i}=0 \,.\label{masterdelta}\ee
This solves the sheet thickness given the inflow and the outflow velocity at an arbitrary point on the boundary. 

In the far outflow region $y\approx L\gg Y$, (\ref{master2}) implies
\be  \fft{\hat{B}_o}{\hat{B}_0}\sim \hat{v}_i^2\fft{y h_1}{\Delta}\sim \hat{v}_i \,,\ee
in agreement with previous estimations. On the other hand, near the diffusion region $y\sim Y\,,\Delta\sim X$, one has 
\be \fft{\hat{B}_o}{\hat{B}_0}\approx \fft{6\hat{v}_i^3}{\eta}yh_1(y) \,.\ee
This implies that in the outflow region, the reconnected magnetic field grows and approaches to a constant asymptotically.

To solve the outflow velocity at $O$, we integrate the momentum equation (\ref{emeqr}) along the radial direction. We deduce
\be h\Big(\hat{\gamma}_o^2 \hat{v}_o^2+\hat{\gamma}_o^2 \hat{v}_o\hat{v}_i\,\fft{y h_1}{\Delta}+\hat{\gamma}_o^2\, y\partial_y\mm{ln}\alpha \Big)=p\Big( 1+\fft{2y h_1}{\Delta}\fft{\hat{B}_o}{\hat{B}_0} \Big) \,.\label{master3}\ee
This, together with (\ref{master1}), (\ref{master2}) (or (\ref{masterdelta})) provides a complete set of equations to solve quantities $\Delta\,,\hat{B}_o\,,\hat{v}_o$ at any position on the outflow-inflow boundary, with a given reconnection rate. It turns out that the situation could be much simplified since only mildly relativistic outflow solution is permitted. To see this, consider the far region $y\approx L$ and using (\ref{deltafar}), (\ref{rate1}), we deduce from (\ref{master3})
\be \big( 1+\hat{\gamma}_o\big)\hat{\gamma_o}^2\hat{v}_o^2+\hat{\gamma}_o^2\, L\partial_y\mm{ln}\alpha\Big|_{y\approx L}\approx\fft{1}{4}\big( 1+2\hat{\gamma}_o \big) \,.\ee
This is a cubic equation for $\hat{\gamma}_o$, of which the general solution is complex. However, clearly an ultra relativistic solution is not allowed. We may take $\hat{\gamma}_o\approx 1$ and then\footnote{Since the tidal force is negligible, one can solve the outflow velocity in terms of small $ L\partial_y \mm{ln}\alpha$ series
\bea 
&&\hat{\gamma}_o\approx 1.176970-0.276516\, L\partial_y \mm{ln}\alpha\Big|_{y\approx L}\,,\nn\\
&&\hat{v}_o\approx 0.527315-0.321662\, L\partial_y \mm{ln}\alpha\Big|_{y\approx L}\,.
\eea}
\bea
&&\hat{v}_o=\Big( \fft{1-L\partial_y \mm{ln}\alpha }{2} \Big)^{1/2}\Big|_{y\approx L}\,,\nn\\
&&\hat{\gamma}_o=\Big( \fft{1+L\partial_y \mm{ln}\alpha }{2} \Big)^{-1/2}\Big|_{y\approx L}\,.
\eea
The results are essentially the same as the Sweet-Parker-like model \cite{Asenjo:2017gsv}. This does not conflict with our setup since the Petschek model improves the reconnection rate rather than the kinetic energy of the outflow significantly.

Finally, to complete our derivations, we compute the reconnection rate, according to the initial condition of the magnetic field in the inflow region. The more the magnetic field line bends down, the larger the reconnection rate becomes. Consider the upper half region $\hat{B}_{\mm{in}}=-\hat{B}_0 e_r+\hat{B}_1$. We assume $|\hat{B}_1|\ll \hat{B}_0$ and the ingoing magnetic field can be well described by a static potential. On the boundary, continuity of magnetic field lines demands 
\be \hat{B}_{\perp}=\hat{B}_{\mm{in}\perp} \,.\ee
The slope of the boundary at $O$ can be specified as
\be \tan{\psi}\equiv\fft{\Delta(y)}{y h_1(y)} \,.\ee 
It follows that $\hat{B}_{\perp}=-\hat{B}_o\cos\psi$ and $\hat{B}_{\mm{in}\perp}=\hat{B}_0
\sin\psi+\hat{B}_{1\perp}$. However, since $|\hat{B}_1|\ll \hat{B}_0$, $\psi$ should be very small ( this is a very good approximation for typical reconnection rate $\hat{v}_i=0.1\sim 0.001\ll 1$). Hence in the far outflow region $\psi\approx \hat{v}_i\approx \hat{B}_o/\hat{B}_0$. One has $\hat{B}_{1\perp}\approx \hat{B}_0\psi-\hat{B}_o\approx  -2\hat{v}_i\hat{B}_0$. In the lower half region, symmetry of the current sheet implies $\hat{B}_{1\perp}\approx 2\hat{v}_i\hat{B}_0$. Therefore, the $\hat{B}_1$ field in the inflow region can be viewed as the static field produced by two monopoles $\hat{B}_{1\perp}\approx \mp 2\hat{v}_i\hat{B}_0$. 
Using standard approach, the magnetic field at the center incidence point $i$ can be evaluated as 
\be \hat{B}_1^r\Big|_i\approx \fft{2\hat{v}_i\hat{B}_0}{\pi}\mm{ln}\Big( \fft{L^2}{Y^2}\Big) \approx \fft{4\hat{v}_i\hat{B}_0}{\pi}\mm{ln}\big( \hat{v}_i^2 S_ih_1(r_o)\big) \,,\ee
where in the second equality $L/Y\approx  \hat{v}_i^2 S h_1(r_o)$ if $L\ll r_g$ according to $(\ref{diffusionrate})$. 
\begin{figure}
\centering
\includegraphics[width=140pt]{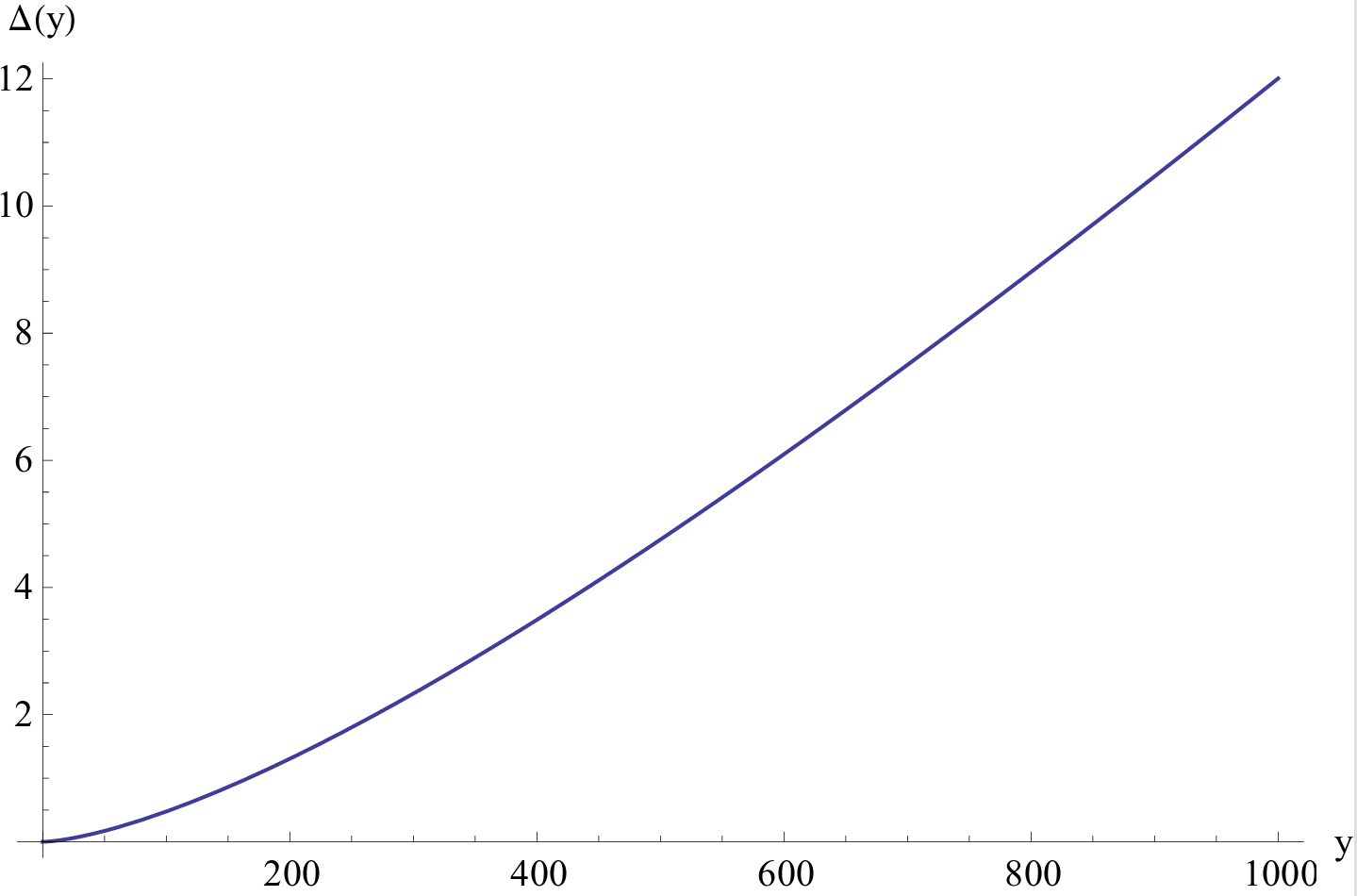}
\includegraphics[width=140pt]{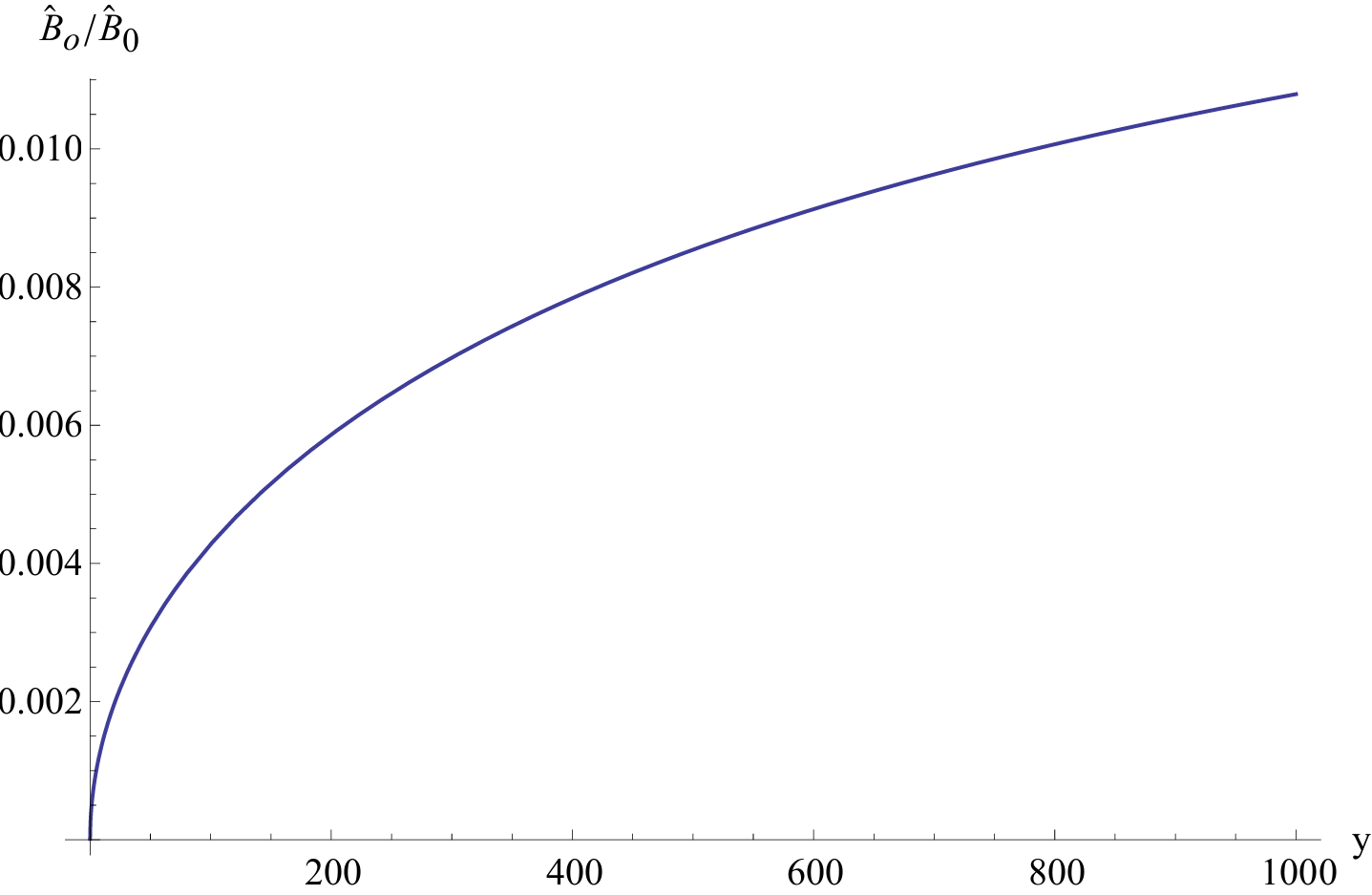}
\includegraphics[width=140pt]{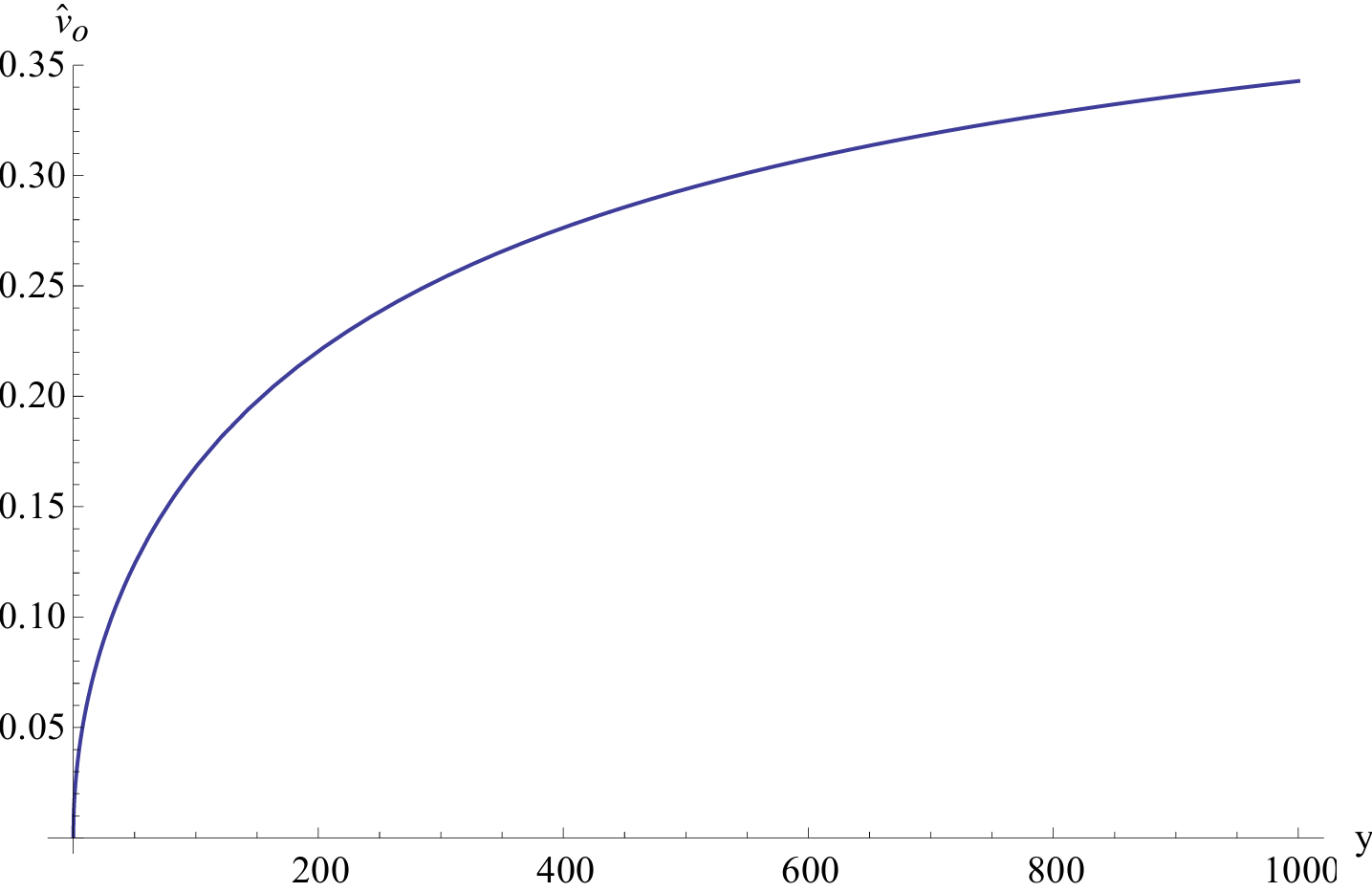}
\caption{Numerical solution for sheet thickness $\Delta$, magnetic field $\hat{B}_o/\hat{B}_0$ and outflow velocity on the inflow-outflow boundary. Here we take $r_C=99r_g/10\,,L=r_g/10\,,k=\pi/8\,,S=10^{18}$, which gives the reconnection rate $\hat{v}_i\approx 0.012$. We have set $a=9/10\,,r_g=10000$.}
\label{bdysol}
\end{figure}
Since $\hat{B}_{in}^r=-\hat{B}_0+\hat{B}_1^r<0$, it implies that the reconnection rate can be significantly improved to $\hat{v}_i\sim (\mm{ln}S)^{-1}$. We deduce
\be \hat{v}_i\approx k\,\Big(\mm{ln}\big(\hat{v}_i^2\,Sh_1(r_o)\big)\Big)^{-1} \,,\qquad k=\fft{\pi\hat{B}_1^r}{4\hat{B}_0} \Big|_i\,.\ee 
Since $\hat{B}_1^r<\hat{B}_0$, a typical choice is $k=\pi/8$ when $\hat{B}_1^r=\hat{B}_0/2$. This tells us that gravity effect decreases the reconnection rate with respect to the flat spacetime limit. This can be seen even more clearly by expanding the result in large $r_o$
\be \hat{v}_i\approx k\Big[\mm{ln}\big(\hat{v}_i^2S\big)+\fft{r_g}{r_o}+\fft{(2-a^2)r_g^2}{2r_o^2}+\cdots \Big]^{-1} \,.\ee

In Fig. \ref{bdysol}, an example of numerical solutions for the sheet thickness $\Delta$, the magnetic field $\hat{B}_o$ and the outflow velocity $\hat{v}_o$ in Kerr black holes is presented. Here we take $r_C=99r_g/10\,,L=r_g/10\,,k=\pi/8\,,S=10^{18}\,,a=9/10\,,r_g=10000$, which results to a reconnection rate $\hat{v}_i\approx 0.012$. We consider mildly relativistic outflow $\hat{\gamma}_o\approx 1$, which is justified by the numerical solution. The length scales of the diffusion region $X\sim 10^{-13}\,,Y\sim 10^{-10}$ are highly suppressed compared to the length of sheet $L=1000$. The shape of the boundary approaches to a straight line with the slope $\sim \hat{v}_i$ rapidly. The reconnected magnetic field increases monotonically and approaches to $\hat{v}_i \hat{B}_0$ asymptotically.

\section{Reconnection layer in azimuthal direction}\label{sec4}

\begin{figure}
\centering
\includegraphics[width=250pt]{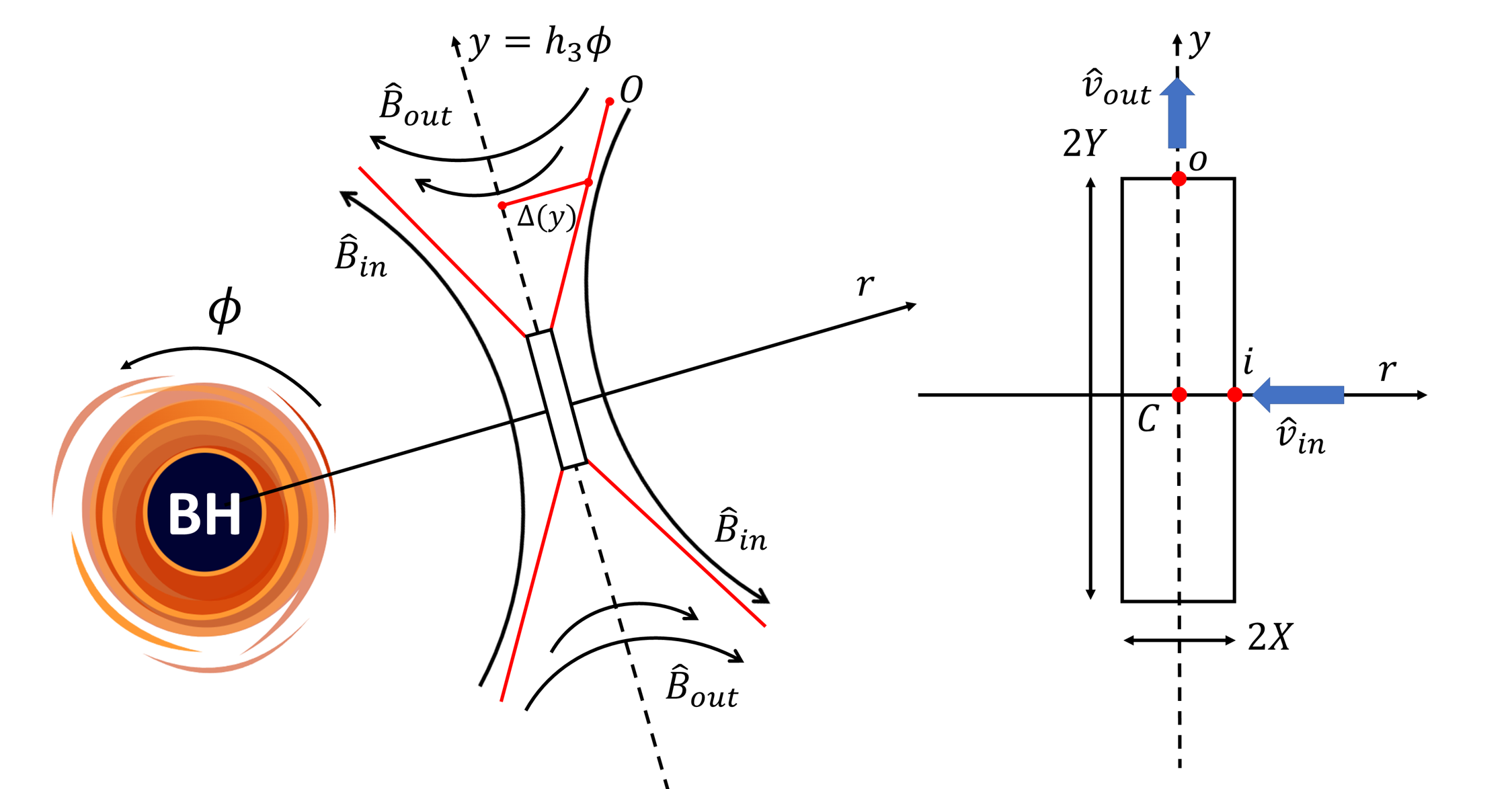}
\caption{Reconnection layer in azimuthal direction. Here $x\equiv r-r_C\,,y\equiv h_3\phi$.}
\label{phipetschek}
\end{figure}

Consider reconnection layer in the azimuthal direction, see Fig. \ref{phipetschek}. Again there are three patches in the current sheet. Without loss of generality, we focus on the upper right region. In the inflow region, the initial magnetic field upstream of the current sheet is specified as: $\hat{B}_{\mm{in}}=-\hat{B}_0 e_{\phi}+\hat{B}_1$ with a small incidence angle $|\hat{B}_1|\ll \hat{B}_0$. In the diffusion region. Sweet-Parker process dominates, with the reconnection rate \cite{Asenjo:2017gsv}
\be \mc{M}\sim \fft{h_1 X}{Y}\Big|_{C}\sim \Big(\fft{Y h_3}{\eta r}\Big)^{-1/2}\Big|_C \,.\label{diffusionratephi}\ee
To study the outflow region, we construct a local orthogonal frame in the $r-\phi$ plane, with $C$ located at the origin. We define $x= r-r_C\,,y=h_3\phi$. The thickness of the current sheet depends on $y$, that is $\Delta=\Delta(y)$ as well as the reconnected magnetic field $\hat{B}^r$ and the outflow velocity $\hat{v_o}$. 

Again we work in the quasi-stationary limit $\partial_t\approx 0$. We assume $\hat{v}^\theta\approx 0 \approx \hat{B}^\theta$ and $\partial_\theta\approx 0$. The current and the electric field are given by $J=\hat{J}^\theta e_\theta\,,E=\hat{E}^\theta e_\theta$ and we take $\hat{v}_{\mm{in}}=-\hat{v}_i e_r$. Consider an arbitrary point $O$ on the boundary. The $\phi$-component of the energy-momentum equation is given by
\be \fft{\partial}{\partial y}\Big(h\hat{\gamma}^2\hat{v}^\phi\big(\hat{v}^\phi+\beta^\phi \big) \Big)+\fft{\partial}{\partial r}\Big(h_1^{-1}h\hat{\gamma}^2 \hat{v}^r \hat{v}^\phi \Big)= -\fft{\partial p}{\partial y}-\hat{J}^\theta\hat{B}^r \,,\label{emeqphi}\ee
where the second term on the l.h.s cannot be dropped since $O$ is located on the boundary. We consider $r_g/r_o\ll 1$ so that $\hat{v}^\phi\gg \beta^\phi\sim r_g^2/r_o^2$, valid to mildly relativistic outflow. Again we focus on a short current sheet in weak gravity regimes so that when integrating the momentum equation, we ignore the contributions from the gravitational tidal force. 

The balance between the inflow flux $\sim\rho\hat{\gamma}_i \hat{v}_i/(h_1\Delta)$ and the outflow flux $\sim \rho\hat{\gamma}_o \hat{v}_o/y$ gives rise to the first estimation of the sheet thickness
\be \Delta(y)\approx \fft{\hat{\gamma}_i\hat{v}_i}{\hat{\gamma}_o\hat{v}_o}\,\fft{y}{h_1} \,.\label{deltaphi1}\ee
The Ohm's law on the current sheet gives
\be \hat{\gamma} \hat{E}^\theta+\hat{\gamma}\hat{v}^\phi\hat{B}^r=\eta \hat{J}^\theta \,.\label{ohmphi}\ee
Again at the center incidence point, the plasmas is viewed as ideal MHD so that the electric field is given by $\hat{E}^\phi|_i\approx \hat{v}_i \hat{B}_0$.  
In addition, the electric field on the current sheet is homogeneous $\hat{E}^\phi_o\approx \hat{E}^\phi|_i$ as long as the thickness of the sheet is very small. The electric current at $O$ can be evaluated from the Maxwell's law
\be \hat{J}^\theta\Big|_o\approx h_1^{-1}\partial_r \hat{F}^{\theta r}\approx \fft{\hat{B}_0}{h_1\Delta(y)} \,.\ee
Combining these results, we deduce
\be \Delta(y)-\fft{\hat{\gamma}_i\hat{B}_o}{\hat{\gamma}_o \hat{B}_0}\,\fft{y}{ h_1}-\fft{\eta}{\hat{\gamma}_o \hat{v}_ih_1}=0 \,,\label{masterphi1}\ee
where we have set $\hat{B}^r\Big|_O=-\hat{B}_o$. This connects the thickness of the current sheet to the reconnected magnetic field on the boundary. Close to the diffusion region, the second term can be dropped so that $\Delta\approx \eta/\hat{v}_ih_1\approx X$, consistent with the Sweet-Parker like model \cite{Asenjo:2017gsv}. Otherwise, in the far outflow region $y\approx L$, the third term can be dropped so that asymptotically
\be \fft{h_1\Delta(y)}{y }\Big|_{y\approx L}\approx \fft{\hat{\gamma}_i \hat{B}_o}{\hat{\gamma}_o \hat{B}_0} \,.\label{deltafar-phi}\ee
Combing the result with (\ref{deltaphi1}), one finds the reconnection rate is related to the asymptotic thickness or the magnetic field
\be \hat{v}_i\sim\fft{\hat{B}_o}{ \hat{B}_0} \approx \fft{h_1\Delta(y)}{y}\,,\label{rate1-phi}\ee
where we have assumed the outflow is mildly relativistic $\hat{\gamma}_o\approx 1$, which will be verified later. It implies that asymptotically the shape of the boundary approaches to a straight line, with the slope $\sim \hat{v}_i$. 

Integrating the momentum equation (\ref{emeqphi}) along the radial direction yields
\be \fft{\hat{B}_o}{\hat{B}_0}=\fft{h\hat{v}_i^2}{\hat{B}_0^2}\Big[ 2y\partial_y\big( \fft{y }{h_1\Delta} \big)+\fft{\hat{\gamma}_o y}{h_1\Delta} \Big]\,,\label{masterphi2}\ee
where we have adopted (\ref{deltaphi1}). Furthermore, by plugging this equation into (\ref{masterphi1}), we deduce
\be h_1\Delta(y)-2\hat{v}_i^2y\Big[2\hat{\gamma}_o^{-1}y\partial_y\big( \fft{y}{h_1\Delta} \big)+\fft{y}{h_1\Delta}\Big]-\fft{\eta}{\hat{\gamma}_o\hat{v}_i}=0 \,.\label{masterdeltaphi}\ee
This determines the shape of the current sheet given the plasmas velocities $\hat{v}_i$ and $\hat{v}_o$. It implies in the far outflow region $y\approx L$, $\Delta(y)\sim\hat{v}_iy/h_1$ so that
\be  \fft{\hat{B}_o}{\hat{B}_0}\sim \hat{v}_i^2\fft{y}{h_1\Delta}\sim \hat{v}_i \,,\ee
in agreement with previous estimations. On the other hand, near the diffusion region $y\sim Y\,,\Delta\sim X$, one finds instead
\be \fft{\hat{B}_o}{\hat{B}_0}\approx \fft{6\hat{v}_i^3}{\eta}y \,.\ee
These results imply that in the outflow region, the reconnected magnetic field grows monotonically and approaches to a constant asymptotically.

To solve the outflow velocity at $O$, we integrate the momentum equation (\ref{emeqphi}) along the azimuthal direction. We deduce
\be h\Big(\hat{\gamma}_o^2 \hat{v}_o^2+\hat{\gamma}_o^2 \hat{v}_o\hat{v}_i\,\fft{y }{h_1\Delta}\Big)=p\Big( 1+\fft{2y}{h_1\Delta}\fft{\hat{B}_o}{\hat{B}_0} \Big) \,.\label{masterphi3}\ee
This, together with (\ref{masterphi1}), (\ref{masterphi2}) or (\ref{masterdeltaphi}) provides a complete set of equations to solve various quantities $\Delta\,,\hat{B}_o\,,\hat{v}_o$ at an arbirtary point on the inflow-outflow boundary. It turns out that only mildly relativistic outflow solution is permitted. To see this, consider the far outflow region, one has
\be \Big(\hat{\gamma_o}^2\hat{v}_o^2+\hat{\gamma}_o^2\hat{v}_o\Big)\Big|_{y\approx L}\approx\fft{3}{4} \,,\ee
which gives $\hat{\gamma}_o\hat{v}_o\Big|_{y\approx L}\approx 1/2$.

Finally, to complete our derivations, we consider the inflow region and derive the reconnection rate, according to the initial condition of the magnetic field. Again we assume the magnetic field can be well described by a static potential. 

Then standard analysis implies that the $\hat{B}_1$ field in the inflow region can be viewed as the static field produced by two monopoles with $\hat{B}_{1\perp}\approx \mp 2\hat{v}_i\hat{B}_0$. Evaluating the magnetic field at the incidence point $i$ yields
\be \hat{B}_1^\phi\Big|_i\approx \fft{2\hat{v}_i\hat{B}_0}{\pi}\mm{ln}\Big( \fft{L^2}{Y^2}\Big)\approx \fft{4\hat{v}_i\hat{B}_0}{\pi}\mm{ln}\big( \hat{v}_i^2 Sh_3(r_o)/r_o\big)  \,,\ee
where in the second equality we have adopted $L/Y\approx \hat{v}_i^2 S h_3(r_o)/r_o$ if $L\ll r_g$.
\begin{figure}
\centering
\includegraphics[width=140pt]{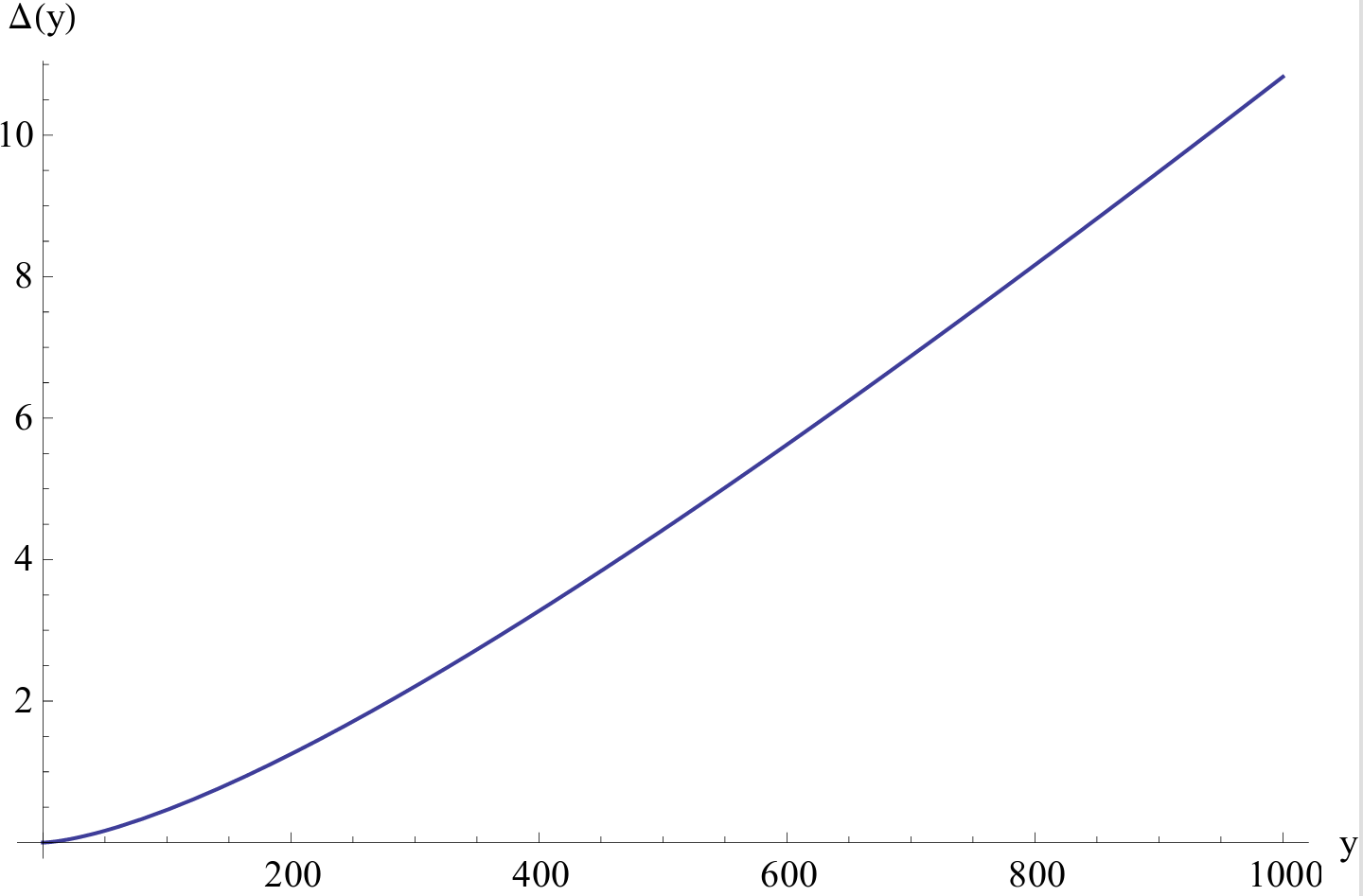}
\includegraphics[width=140pt]{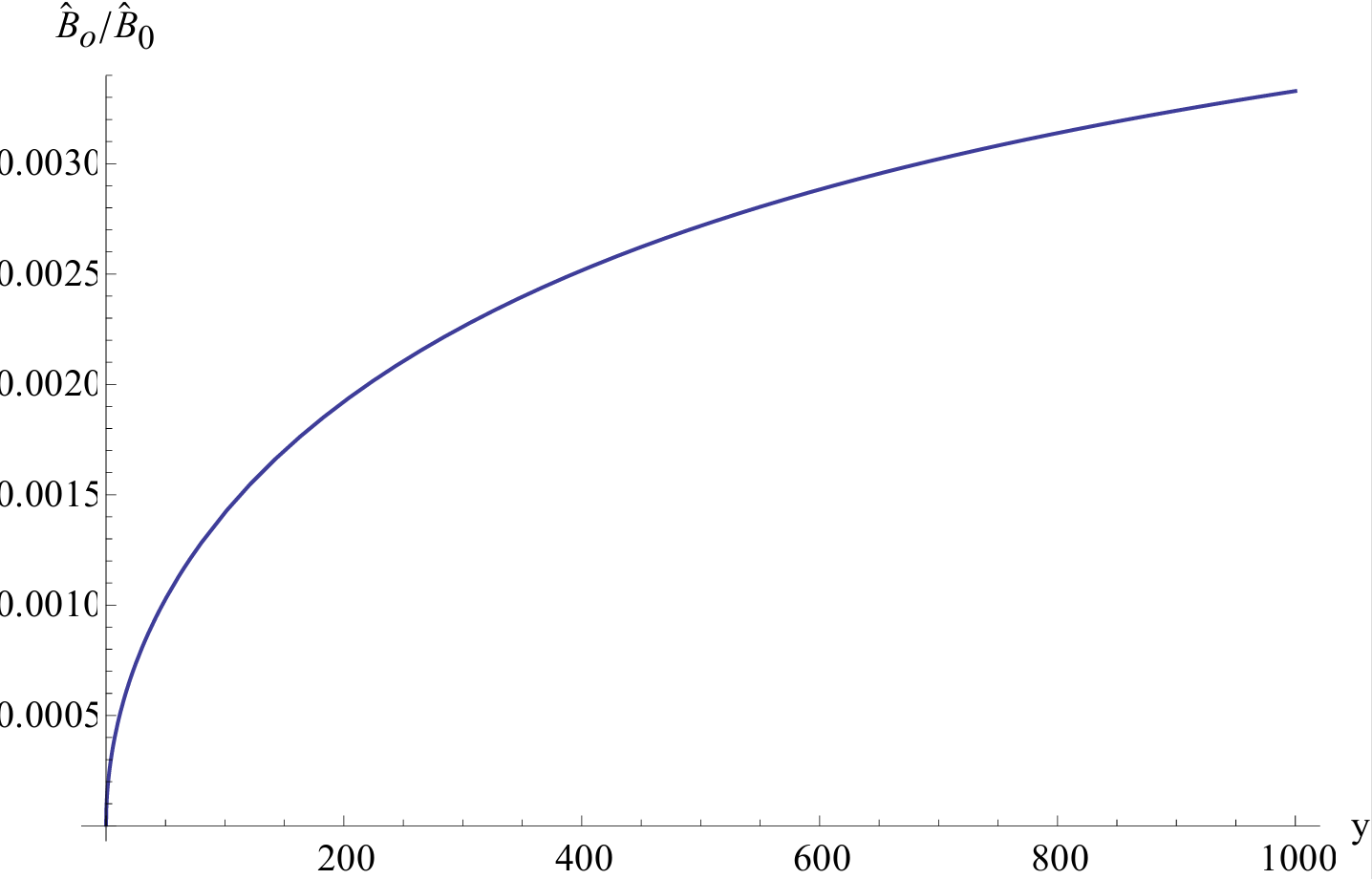}
\includegraphics[width=140pt]{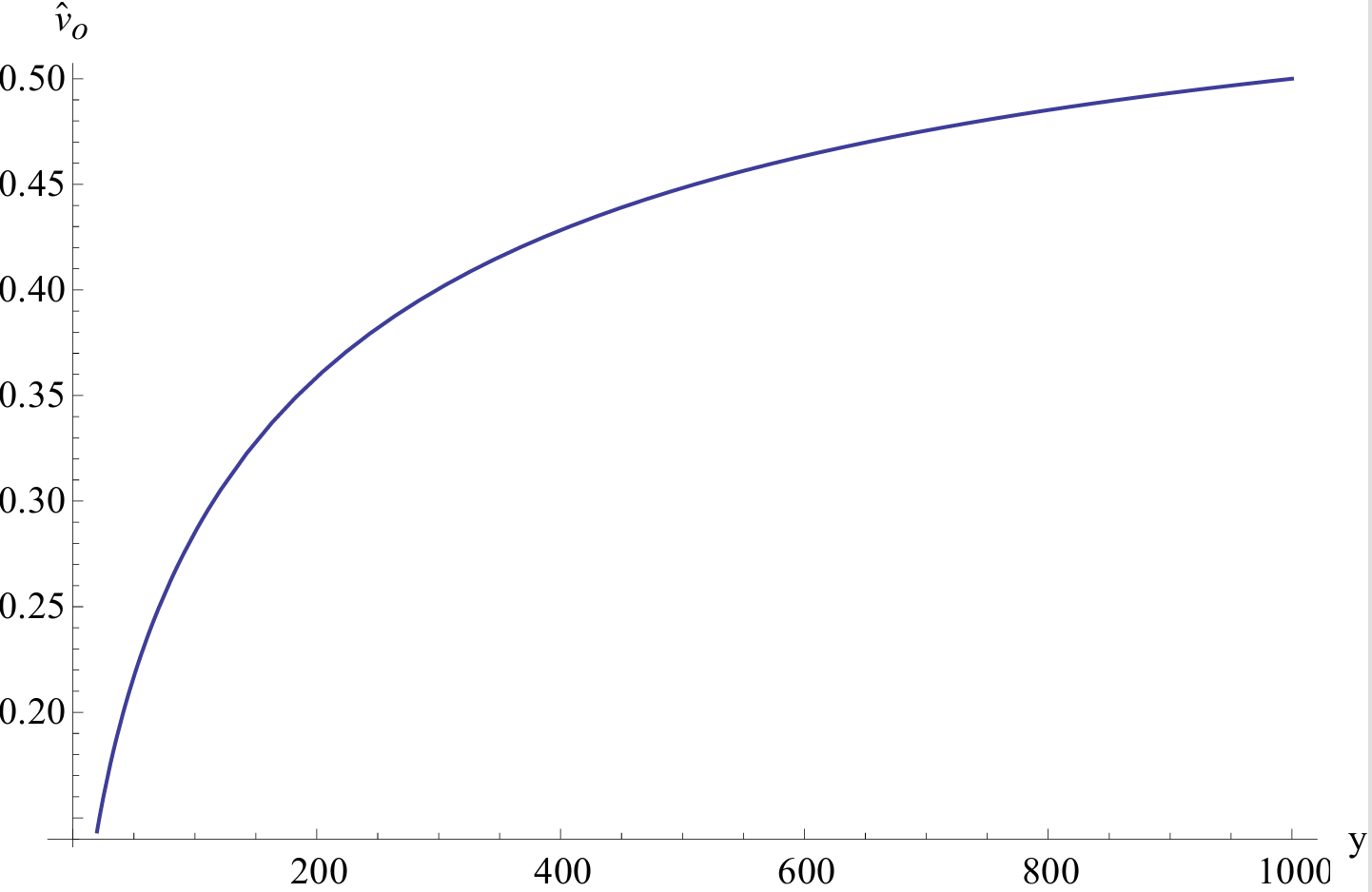}
\caption{Azimuthal reconnection layer: numerical solutions for sheet thickness $\Delta$, magnetic field $\hat{B}_o/\hat{B}_0$ and outflow velocity on the inflow-outflow boundary. Here we take $r_C=99r_g/10\,,L=r_g/10\,,k=\pi/8\,,S=10^{18}$ and $h_1$ is evaluated at $r=10r_g$. This results to the reconnection rate $\hat{v}_i\approx 0.012$. We have set $a=9/10\,,r_g=10000$.}
\label{bdysolphi}
\end{figure}
Again the reconnection rate can be significantly improved to $\hat{v}_i\sim (\mm{ln}S)^{-1}$. One has
\be \hat{v}_i\approx k\,\Big(\mm{ln}\big(\hat{v}_i^2\,Sh_3(r_o)/r_o\big)\Big)^{-1} \,,\qquad k=\fft{\pi\hat{B}_1^\phi}{4\hat{B}_0} \Big|_i\,.\ee 
Again the gravity effect decreases the reconnection rate with respect to the flat spacetime limit. The leading corrections is given by 
\be \hat{v}_i\approx k\Big[\mm{ln}\big(\hat{v}_i^2S\big)+\fft{a^2 r_g^2}{2r_o^2}+\cdots \Big]^{-1} \,.\ee 
Unlike the radial case, rotation of the black hole dominates in the curvature effects.

In Fig. \ref{bdysolphi}, we present numerical solutions for the sheet thickness $\Delta$, the magnetic field $\hat{B}_o/\hat{B}_0$ and the outflow velocity $\hat{v}_o$. To compare with the radial case, we choose the same parameters for the plasmas as well as the background. This gives the reconnection rate $\hat{v}_i\approx 0.012$. Clearly all the results are consistent with previous analytical analysis.

\section{Reconnection layer in non-ZAMO frame}\label{sec5}

In curved spacetime, the motion of plasmas is generally complicated and the electric current sheet may rotate either faster or slower than the black hole. Here we would like to explore the effect of rotation of the reconnection layer on the magnetic reconnection from the view of a ZAMO observer. The relevant topic in Minkowski spacetime is the magnetic reconnection for a rotating current sheet  
in the laboratory frame. 

Again we work in quasi-stationarity and consider circular stable orbits. We consider a typical example at first: the Einstein's rotating disk, which orbits around the ZAMO frame with a constant angular velocity $\omega$. General rotating disk will be discussed later. 

Before introducing the rotating disk frame, we first rewrite the Kerr metric in cylindrical coordinates $(\hat{t}\,,\hat{r}\,,\hat{\psi},\hat{z})$ for a ZAMO observer
\be ds^2=-d\hat{t}^2+d\hat{r}^2+\hat{r}^2 d\hat\psi^2+d\hat{z}^2 \,,\ee
where $\hat{x}^1=\hat{r}\cos\hat{\psi}\,,\hat{x}^3=\hat{r}\sin\hat\psi\,,\hat{z}=\hat{x}^2$. Notice that these coordinates are not adapted to local measures. Vectors (or tensors) measured in the local rest frame (LRF) are related to those in the cylindrical coordinates via vielbeins\footnote{For example the metric measured by a local observer is always Minkowski, which is related to the metric in curved spacetime as $\eta_{ab}=g_{\mu\nu}\, e^\mu_{\,\,\,a} e^\nu_{\,\,\,b} \,,$ where $e^\mu_{\,\,\,a}$ stand for the vielbeins. }. It will be convenient for us to work directly with a new coordinate $d\hat{y}=\hat{r}d\hat\psi$, which absorbs the warped factor in the polar direction. This is equivalent to building a orthonormal frame for local observers under cylindrical coordinates.

To proceed, we introduce the rotating disk frame $\{\bar{t}\,,\bar{r}\,,\bar{\psi}\,,\bar{z} \}$, defined as
\be d\bar{t}=d\hat{t}\,,\quad d\bar{r}=d\hat{r}\,,\quad d\bar{\psi}=d\hat{\psi}-\omega d\hat{t}\,,\quad d\bar{z}=d\hat{z} \,.\label{einsteindiskframe}\ee
In this frame, the local metric becomes hypersurface non-orthogonal
\be ds^2=-(1-\omega^2 \bar{r}^2)d\bar{t}^2+d\bar{r}^2+\bar{r}^2d\bar{\psi}^2+2\omega\bar{r}^2 d\bar{t}d\bar{\psi}+d\bar{z}^2 \,.\ee
However, it will be cumbersome to directly study a rotating reconnection layer for the ZAMO observer. The existence of the mixing metric component $\bar{g}_{t\psi}$ will lead to various issues. Our strategy is examining the properties of the reconnection layer for a comoving observer at first and then translating the results into the ZAMO frame. We define the corotating frame (CRF) $\{\tilde{t}\,,\tilde{r}\,,\tilde{y}\,,\tilde{z}\}$ to be
\be d\tilde{t}=d\bar{t}-\fft{\omega\bar{r}^2}{1-\omega^2\bar{r}^2}\,d\bar{\psi}\,,\quad d\tilde{r}=d\bar{r}\,,\quad d\tilde{y}=\bar{r}d\bar{\psi}\,,\quad d\tilde{z}=d\bar{z}  \,,\ee
in which the metric becomes hypersurface orthogonal
\be\label{CRDF} ds^2=-q^2 d\tilde{t}^2+d\tilde{r}^2+q^{-2}d\tilde{y}^2+d\tilde{z}^2  \,,\ee
where $q=\sqrt{1-\omega^2\tilde{r}^2}$. Notice that the spatial metric becomes non-Euclidean for the comoving observers. This is a characteristic feature for a general rotating disk. It is extremely important since measure of (current sheet) length is essentially not local and will effect the reconnection rate significantly. Without loss of generality, we put the frame at $\hat\psi=0$ so that $\hat{r}$ (or $\tilde r$) is parallel to radial direction of the black hole. Vectors or tensors measured by a ZAMO observer are related to those in the CRF via the coordinate transformations
\bea\label{bridge}
&&d\hat t=d\tilde t+\fft{\omega \tilde r}{1-\omega^2\tilde{r}^2} d\tilde{y}\,,\nn\\
&&d\hat{x}^1=d\tilde r\,,\quad d\hat{x}^2=d\tilde z\,,\nn\\
&& d\hat{x}^3=\omega \tilde r d\tilde t+\fft{1}{1-\omega^2\tilde{r}^2}d\tilde y\,.
\eea

\subsection{Sweet-Parker model}
Firstly, we consider a Sweet-Parker model, with the current sheet located in the $\tilde{r}$-direction. We assume $\tilde{v}^z=0=\tilde{B}^z$ and $\tilde\partial_{z}\approx 0$. The electric and magnetic fields are related to the filed strength tensor as\footnote{The covariant definitions are $\tilde{E}_\mu=\tilde{F}_{\mu\nu}\tilde{\xi}^\nu\,,\tilde{B}_\mu=-{}^*\tilde{F}_{\mu\nu}\tilde{\xi}^\nu$, where $\tilde{\xi}$ is the timelike Killing vector and ${}^*\tilde{F}$ is the dual field strength.}
\be \tilde{E}_i=-q^{-1}\tilde{F}_{0i}\,,\quad \tilde{B}_i=\fft 12q^{-1} \epsilon_{ijk}\tilde{F}^{jk}  \,.\ee
\begin{figure}
\centering
\includegraphics[width=300pt]{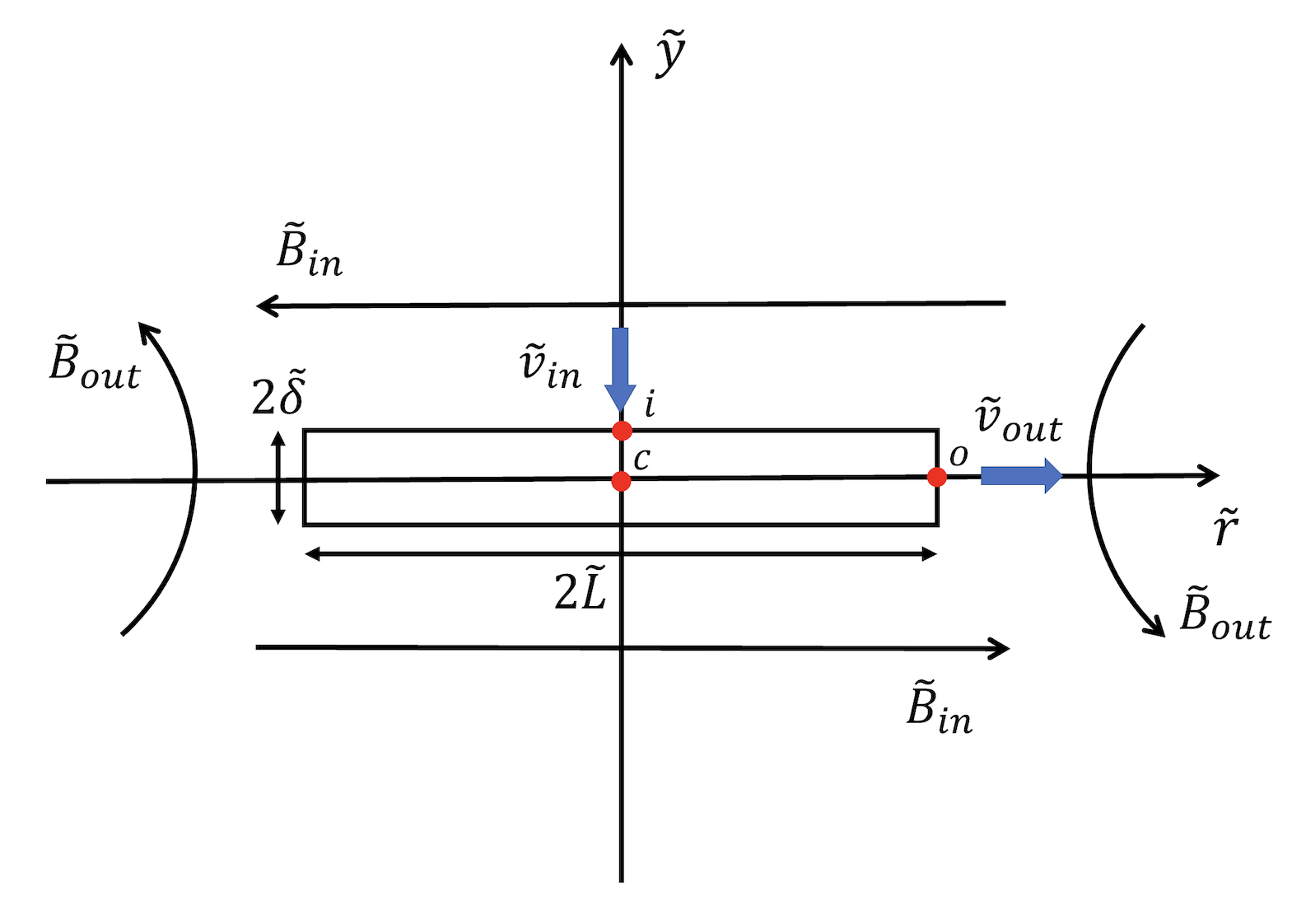}
\caption{Radial magnetic reconnection layer of Sweet-Parker configuration in corotating frame.}
\label{radialdisk}
\end{figure}
Here the $q^{-1}$ factor for the electric field comes from the timelike Killing vector whereas the same factor for the magnetic field comes from the three volume of the spatial slice. Without loss of generality, we focus on the first quadrant of  Fig. \ref{radialdisk}, where we set $\tilde{B}_{\mm{in}}=-\tilde{B}_0 \tilde e_r\,,\tilde{v}_{\mm{in}}=-\tilde{v}_i \tilde e_y$ and $\tilde{B}_{\mm{out}}=-\tilde{B}_o \tilde e_y\,,\tilde{v}_{\mm{out}}=\tilde{v}_o \tilde e_r$.
From continuity equation, the inflow flux $\tilde\partial_{y}(\tilde{\rho}\tilde{\gamma}_i\tilde{v}_i)\approx \fft{\tilde{\rho}\tilde{\gamma}_i\tilde{v}_i}{\tilde\delta} $ must balance the outflow flux $\tilde\partial_{r}(\tilde{\rho}\tilde{\gamma}_o\tilde{v}_o)\approx \fft{\tilde{\rho}\tilde{\gamma}_o\tilde{v}_o}{\tilde L}$. This gives
\be \tilde\delta=\fft{\tilde{\gamma}_i\tilde{v}_i}{\tilde{\gamma}_o \tilde{v}_o}\,\tilde{L}  \,.\label{deltadisk}\ee 
Here $\tilde{\gamma}$ is the Lorentz factor in the CRF.   Normalization of four-velocity requires $\tilde{\gamma}_i=(1-\omega^2\tilde{r}^2-q^{-2}\tilde{v}_i^2)^{-1/2}$ and $\tilde{\gamma}_o=(1-\omega^2\tilde{r}^2-\tilde{v}_o^2)^{-1/2}$. Notice that reality of the Lorentz factor requires $q\geq q^{-1}\tilde{v}_i$ and $q\geq \tilde{v}_o$. This could be understood as the disk is slowly rotating compared to the plasmas velocities. In fact, this is a consequence of causality, as will be shown later. According to the Ohm's law, evaluation of the electric field at i and o  yields
\bea
&& E_z|_i=q^{-2}\tilde{v}_i\tilde{B}_0 \,,\nn\\
&& E_z|_o=q^{-2}\tilde{v}_o \tilde{B}_o \,.
\eea
Here and below $q$ (and $\tilde{r}$) is understood to be evaluated at $O$. Again we assume the electric field on the current sheet is uniform so that  
\be \tilde{B}_o=\fft{\tilde{v}_i \tilde{B}_0}{\tilde{v}_o}\approx  \fft{\tilde{\delta}\tilde{B}_0 }{\tilde{L}} \,,\label{magneticdisk}\ee
for mildly relativistic outflow $\tilde{\gamma}_o\approx 1$. The result is in agreement with magnetic flux conservation.

To derive the outflow velocity, we consider the energy-momentum equation in $\tilde{r}$-direction. To leading order in large $\tilde{r}$, one has
\be\label{emr-disk} \partial_{\tilde r}\big( h\tilde{\gamma}^2\tilde{v}^2\big)\approx -(\partial_{\tilde r}p+q\tilde{J}^z \tilde{B}_y) \,.\ee
On the other hand, according to Maxwell's equation
\be \tilde{J}^z\Big|_O\approx\tilde\partial_{y}\tilde{F}^{zy}\Big|_O\approx \fft{q\tilde{B}_0}{\tilde\delta} \,.\ee
Integrating the equation (\ref{emr-disk}) yields $h\tilde{\gamma}_o^2\tilde{v}_o^2\approx p+\tilde{B}_0^2$. This gives 
\be \tilde{\gamma}_o\tilde{v}_o\approx 1 \,.\ee
Indeed the outflow is mildly relativistic. To study the reconnection rate, consider the Ohm's law on the current sheet
\be q\tilde{\gamma}(\tilde{E}_z+\tilde{v}\tilde{B}_y  )=\eta\tilde{J}^z \,.\ee
On the neutral line $\tilde{v}=0$ so that $ \tilde{E}_z|_C\approx \eta \tilde{J}^z/q $. This leads to 
\be \tilde{J}^z\Big|_C\approx \fft{\tilde{v}_i \tilde{B}_0}{q\eta } \,.\ee
Since $\tilde{J}^z|_C\approx \tilde{J}^z|_O$ in our approximation, one finds $\tilde{v}_i=q^2\eta/\tilde{\delta}$. It turns out that
\be
\tilde{v}_i=q\,u_i\,,\quad u_{i}=h_{1}^{-1/2} \Big|_o\,\,S^{-1/2}\,,
\label{ratedisk}\ee
where $u_{ i}$ is the reconnection rate locally measured by the comoving observer, which is irrelevant to motion of the plasmas. In contrast, the reconnection rate for the ZAMO observer does rely on rotation of the plasmas. Using (\ref{bridge}) and $\hat{U}^\mu=\tilde{U}^\nu \partial\hat{x}^\mu/\partial\tilde{x}^\nu$, we compute the four-velocity at i:
$\hat{U}^0=\tilde{\gamma}_i(1-q^{-2}\tilde{v}_{i}\,\omega \tilde{r})$ and $\hat{U}^3=\tilde{\gamma}_i(\omega\tilde r-q^{-2}\tilde{v}_{i})$. This gives rise to
\be \hat{v}_i\equiv -\hat{U}^3/\hat{U}^0= \fft{q^{-2}\tilde{v}_{i}-\omega \tilde r}{1-q^{-2}\tilde{v}_{i}\,\omega \tilde{r}}\Big|_o \,.\label{ratezamo}\ee
Clearly the result does not follow from the speed superposition principle in special relativity owing to the warped factor $q^{-2}$. This can be traced back to the fact that the spatial space in CRF is non-Euclidean. Nevertheless, causality is protected since $\hat{v}_i\leq 1$ because of $u_i\leq q$, which guarantees reality of the Lorentz factor $\tilde{\gamma}_i$. 
\begin{figure}
\centering
\includegraphics[width=190pt]{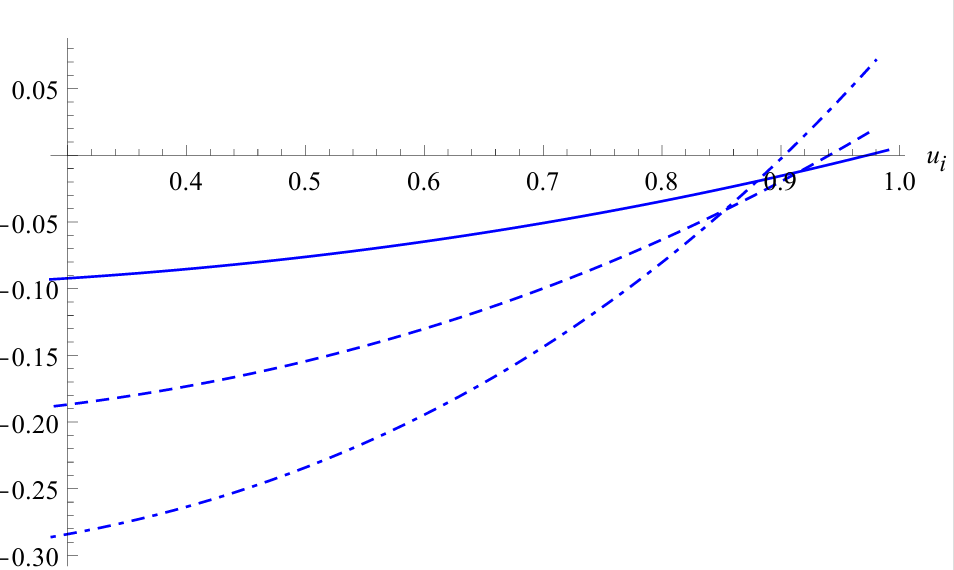}
\includegraphics[width=190pt]{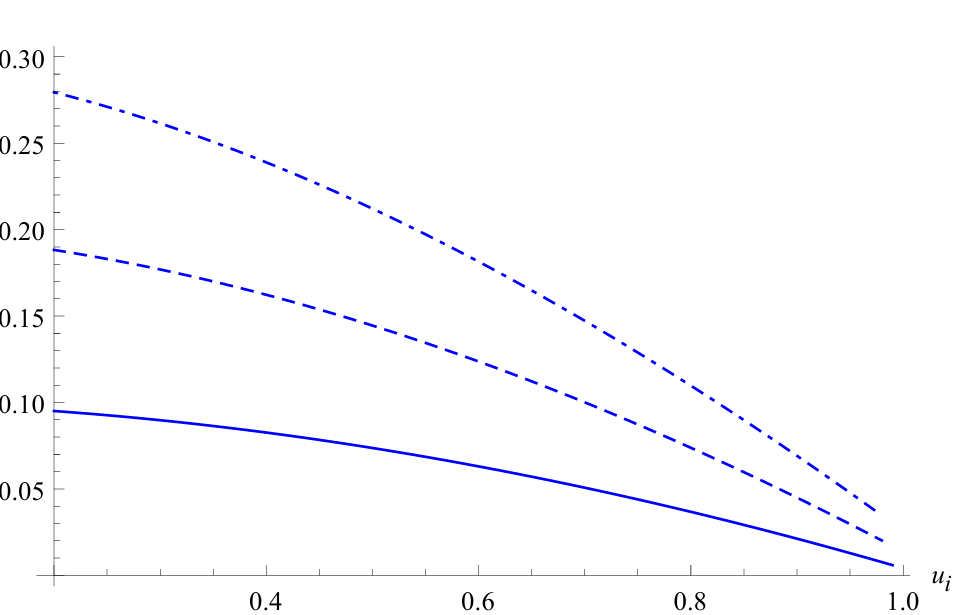}
\caption{Plots for the difference $\hat{v}_i-u_i$ as a function of $u_i$. In the left panel $\omega>0$ and the solid/dashed/dotdashed line corresponds to $\omega\tilde{r}=0.1/0.2/0.3$, respectively. In the right panel $\omega<0$ and the solid/dashed/dotdashed line corresponds to $\omega\tilde{r}=-0.1/-0.2/-0.3$, respectively.}
\label{raterotation}
\end{figure}
It turns out that in comparison to the rate $u_i$ for the comoving observer, the reconnection rate $\hat{v}_i$ measured by the ZAMO observer could be either decreased or increased depending on rotation of the disk. There exists a critical value of $u_i$
\be u_c^{\pm}=\fft{-(1-q)\pm\sqrt{(1-q)^2+4q\omega^2\tilde{r}^2}}{2\omega\tilde{r}} \,,\ee
where $u_c^+$ ($u_c^-$) corresponds to $\omega>0$ ($\omega<0$). Consider $\omega>0$ at first. In this case, the disk rotates slower than the ZAMO frame. One has $u_c^-<0\,,u_c^+\leq 1$ and $u_c^+$ decreases as $\omega\tilde r$ increases. Since $\hat{v}_i-u_i\sim \omega(u_i-u_c^+)(u_i-u_c^-)$, it follows that $\hat{v}_i<u_i$ if $u_i<u_c^+$ and $\hat{v}_i>u_i$ if $u_i>u_c^+$. However, for subrelativistic inflow, the reconnection rate is generally decreased, see the left panel of Fig. \ref{raterotation}. This can be seen even more clearly for a non-relativistic disk which has $\omega\tilde{r}\sim 0$ and $u_c^+\approx 1-\omega\tilde{r}/4$ approaching to the speed of light. Hence for mildly relativistic inflow (for example for $\omega\tilde r\leq 0.5$, $u_i\leq 0.8\leq u_c^+$), rotation of the reconnection layer will lead to a further decrease of the reconnection rate for the ZAMO observer. 

However, when the disk rotates faster than the ZAMO frame $\omega<0$, the situation will be quite different. The critical velocity is always superluminal $u_c^->1$ for a timelike disk. Since $u_c^+<0$ in this case, it implies that $\hat{v}_i$ is always larger than $u_i$, see the right panel of Fig. \ref{raterotation}.

By symmetry of the configuration, the inflow velocity in the lower half plane of CRF is given by $\tilde{v}_{\mm{in}}=\tilde{v}_i \tilde{e}_y$, where $\tilde{v}_i$ is given by (\ref{ratedisk}). However, in the ZAMO frame, the reconnection rate is obtained by (\ref{ratezamo}), with $\omega\rightarrow -\omega$. Consequently, for any given $\omega\neq 0$, the ZAMO observer will always find the asymmetric rates: one side of the current sheet has a larger reconnection rate whereas the other side has a smaller one, in comparison to the unrotation limit.  This is correct for mildly relativistic inflow no matter the reconnection layer rotates faster or slower than the ZAMO frame.

However, when the current sheet forms in azimuthal direction, the reconnection rate for the ZAMO observer will not receive any contributions from the rotation. We refer the readers to the appendix for details about this case.

\subsection{Petschek model}
We proceed to study a general rotating reconnection layer in the Petschek model, located in the radial direction (the illustration figure is the same as Fig. \ref{radialpetschek} but now all quantities have a tilde). Again we assume $\tilde{v}^z=0=\tilde{B}^z$ and $\tilde\partial_{z}\approx 0\approx \tilde\partial_{t}$. We focus on the first quadrant. We set $\tilde{B}_{\mm{in}}=-\tilde{B}_0 \tilde e_r+\tilde{B}_1\,,\tilde{v}_{\mm{in}}=-\tilde{v}_i \tilde e_y$ and $\tilde{B}_{\mm{out}}=-\tilde{B}_o \tilde e_y\,,\tilde{v}_{\mm{out}}=\tilde{v}_o \tilde e_r$, where O is an arbitrary point on the inflow-outflow boundary. According to previous discussions, in the diffusion region, Sweet-Parker process dominates and 
\be \tilde{v}_i\sim \fft{\tilde{X}}{\tilde{Y}}\sim q\Big(\fft{\tilde{Y}}{\eta}\Big)^{-1/2} \,.\label{disk-diffusion}\ee
To estimate thickness of the current sheet at O, we evaluate the inflow flux $\tilde\partial_{y}(\tilde{\rho}\tilde{U}^{y})\sim \tilde{\rho}\tilde{\gamma}_i\tilde{v}_i/\Delta(\tilde{r})$ and the outflow flux $\tilde\partial_{r}(\tilde{\rho}\tilde{U}^{r})\sim \tilde{\rho}\tilde{\gamma}_o\tilde{v}_o/\tilde{r}$. The balance between the two implies that
\be \Delta(\tilde{r})=\fft{\tilde{\gamma}_i\tilde{v}_i}{\tilde{\gamma}_o\tilde{v}_o}\tilde{r}\Big|_O \,.\label{disk-delta}\ee
On the current sheet, the Ohm's law gives
\be q\tilde{\gamma}(\tilde{E}_z+\tilde{v}_r \tilde{B}_y)=\eta\tilde{J}_z\,. \label{ohm-disk}\ee
We assume the electric field on the current sheet is uniform. Evaluation of the electric field at the incidence point i yields $\tilde{E}_z|_O=\tilde{E}_z|_i=q^{-2}\tilde{v}_i\tilde{B}_0$. Besides, using the Maxwell's law, $\tilde{J}_z|_O\approx \tilde\partial_{y}\tilde{F}^{zy}\approx q\tilde{B}_0/\Delta$.
Then integrating (\ref{ohm-disk}) along $\tilde y$-direction yields
\be \Delta(\tilde{r})-\fft{\tilde{\gamma}_i\tilde{B}_o}{\tilde{\gamma}_o\tilde{B}_0}\tilde{r}-\fft{q^2\eta }{\tilde{\gamma}_o \tilde{v}_i}=0\,,\label{disk-master1} \ee
where we have adopted (\ref{disk-delta}). Close to the diffusion region, the second term is negligible, one has $\tilde{v}_i^2\sim q^2\eta/\tilde{Y}$, consistent with previous analysis. On the other hand, in the far outflow region, the third term is negligible so that $\tilde{v}_i\sim \Delta/\tilde{r}\sim \tilde{B}_o/\tilde{B}_0$. 

To proceed, consider the energy-momentum tensor equation at $\tilde{r}$-direction
\be \fft{\partial}{\partial \tilde{r}}\Big(h\tilde{\gamma}^2\big(\tilde{v}^r \big)^2 \Big)+\fft{\partial}{\partial \tilde{y}}\Big(h\tilde{\gamma}^2 \tilde{v}^r \tilde{v}^y \Big)\approx -\fft{\partial p}{\partial \tilde{r}}-q\tilde{J}^z\tilde{B}_
y \,,\label{disk-emeqr}\ee
where we have dropped metric derivative terms which are subdominant in large $\tilde r$. Integrating (\ref{disk-emeqr}) along $\tilde{y}$-direction and using (\ref{disk-delta}), we deduce
\be \fft{\tilde{B}_o}{\tilde{B}_0}=\fft{h\tilde{v}_i^2}{\tilde{B}_0^2}\Big( 2\tilde{r}\partial_{\tilde{r}}\big( \fft{\tilde{r}}{\Delta} \big)+\fft{\tilde{\gamma}_o \tilde{r}}{\Delta} \Big) \,.\label{disk-master2} \ee
Substituting this equation into (\ref{disk-master1}), we deduce
\be \Delta(\tilde{r})-2\tilde{v}_i^2 \tilde{r}\Big(2\hat{\gamma}_o^{-1}\tilde{r}\partial_{\tilde{r}}\big( \fft{\tilde{r}}{\Delta} \big)+\fft{\tilde{r}}{\Delta}\Big)-\fft{q^2\eta}{\tilde{\gamma}_o\tilde{v}_i}=0 \,.\label{disk-masterdelta}\ee
Again $h=4p$ for hot relativistic plasmas and $p=\tilde{B}_0^2/2$ according to pressure balance at the neutral line. For mildly relativistic outflow, this equation solves the thickness of the current sheet given the inflow velocity. 

To derive the outflow velocity at $O$, we integrate the momentum equation (\ref{disk-emeqr}) along $\tilde{r}$-direction. We deduce
\be h\Big(\tilde{\gamma}_o^2 \tilde{v}_o^2+\tilde{\gamma}_o^2 \tilde{v}_o\tilde{v}_i\,\fft{\tilde{r}}{\Delta} \Big)=p\Big( 1+\fft{2\tilde{r}}{\Delta}\fft{\tilde{B}_o}{\tilde{B}_0} \Big) \,.\label{disk-master3}\ee
The outflow is indeed mildly relativistic since $\tilde{\gamma}_o\tilde{v}_o\approx 1$ in the far outflow region. 

The equations (\ref{disk-master2}), (\ref{disk-masterdelta}) and (\ref{disk-master3}) give a complete set of equations to solve the various quantities in the outflow region once the reconnection rate is given. The later depends on the initial condition of the magnetic field in the inflow region $\tilde{B}_{\mm{in}}=-\tilde{B}_0 \tilde e_r+\tilde{B}_1$. Again we assume $|\tilde{B}_1|\ll \tilde{B}_0$ and the ingoing magnetic field can be well described by a static potential.  Derivation of $\tilde{B}_{1}$ can follow the standard approach. One has $\tilde{B}_{1\perp}\approx  -2q^{-1}\tilde{v}_i\tilde{B}_0$. This implies that the $\tilde{B}_1$ field in the inflow region can be viewed as the static field produced by two monopoles $\tilde{B}_{1\perp}\approx \mp 2q^{-1}\tilde{v}_i\tilde{B}_0$. The magnetic field at the center incidence point $i$ can be evaluated as 
\be \tilde{B}_1^r\Big|_i\approx \fft{2\tilde{v}_i\tilde{B}_0}{q\pi}\mm{ln}\Big( \fft{\tilde{L}^2}{\tilde{Y}^2}\Big) \approx \fft{4\tilde{v}_i\tilde{B}_0}{q\pi}\mm{ln}\big( q^{-2}\tilde{v}_i^2 Sh_1(r_o)\big) \,,\ee
where in the second equality $\tilde{L}/\tilde{Y}\approx  q^{-2}\tilde{v}_i^2 S h_1(r_o)$. We deduce
\be \tilde{v}_i\approx q k\,\Big(\mm{ln}\big(q^{-2}\tilde{v}_i^2\,Sh_1(r_o)\big)\Big)^{-1} \,,\quad k=\fft{\pi \tilde{B}_1^r}{4\tilde{B}_0}\,.\label{pratedisk}\ee 
Again $\tilde{v}_i=q u_i$, where $u_i$ is the reconnection rate measured by the comoving observer locally. Similar to the Sweet-Parker case, the rate $\hat{v}_i$ for the ZAMO observer is given by (\ref{ratezamo}). It implies that despite the configuration of current sheet is quite different for the two cases, the reconnection rate is changed in the same manner by rotation of the disk.

\subsection{General rotating disk}
Having studied the Einstein's rotating disk, we would like to relax the condition and explore the reconnection layer in a general rotating frame. Perhaps the simplest way to do this is letting $\omega$ in (\ref{einsteindiskframe}) to be position dependent. While this is correct, here we would like to work in the other way around. Without introducing cylindrical coordinates, we directly define the rotating frame using the BL coordinates as 
\be d\bar{t}=\alpha dt\,,\quad d\bar{x}^i=h_i dx^i-\alpha\bar{\beta}^i dt \,,\ee
where $\bar{\beta}^i=\bar{\beta}^\phi \delta^{i\phi}$. Generally  $\bar{\beta}^\phi\neq \beta^\phi$ and hence this frame does not corotate with the black hole. It is easy to see that this frame is related to the ZAMO frame as
\be d\bar {t}=d\hat{t}\,,\quad d\bar{x}^1=d\hat{x}^1\,,\quad d\bar{x}^2=d\hat{x}^2\,,\quad d\bar{x}^3=d\hat{x}^3-\varpi d\hat{t} \,,\ee
where $\varpi\equiv \bar{\beta}^\phi-\beta^\phi$. The metric turns out to be
\be ds^2=-d\bar{t}^2+(d\bar{x}^1)^2+(d\bar{x}^2)^2+(d\bar{x}^3+\varpi d\bar{t})^2 \,,\ee
which is hypersurface non-orthogonal. To compare with the previous case, we would like to relabel the coordinates properly and define the comoving observer as
\be d\tilde{t}=d\bar t-\fft{\varpi}{1-\varpi^2}d\bar{x}^3\,,\quad d\tilde{r}=d\bar{x}^1\,,\quad d\tilde{y}=d\bar{x}^3\,,\quad d\tilde{z}=d\bar{x}^2 \,.\ee
In this frame, the metric becomes hypersurface orthogonal
\be ds^2=-\mathbf{q}^2 d\tilde{t}^2+d\tilde{r}^2+\mathbf{q}^{-2}d\tilde{y}^2+d\tilde{z}^2 \,,\ee
where $\mathbf{q}=\sqrt{1-\varpi^2}$. Notice that $|\varpi|<1$ for a timelike disk. Clearly when $\varpi=\omega \tilde r$, it reduces to the Einstein's rotating disk. This is consistent with our intuitive expectations that a general rotating disk can be obtained by letting $\omega$ in (\ref{einsteindiskframe}) to be position dependent. 

It is immediately seen that for reconnection layer in radial direction, the inflow velocity $\tilde{v}_i$ will acquire a factor of $\mathbf{q}$ according to (\ref{ratedisk}) or (\ref{pratedisk}) for Sweet-Parker or Petschek scenario. That is $\tilde{v}_i=q u_i$, where the meaning of $u_i$ is the same as before. It turns out that the reconnection rate $\hat{v}_i$ for the ZAMO observer reads
\be \hat{v}_i = \fft{\mathbf{q}^{-2}\tilde{v}_{i}\mp\varpi}{1\mp\mathbf{q}^{-2}\tilde{v}_{i}\,\varpi}\Big|_o \,,\label{ratezamogene}\ee
where ``+'' (``-'') corresponds to the lower/upper half plane of the current sheet. Without presenting more details, we claim that for a generally slowly rotating disk (subrelativistic at most), the reconnection rate $\hat{v}_i$ will be decreased by the rotation on one side of the configuration and be increased on the other side, provided the inflow is mildly relativistic.

\section{Conclusion}\label{sec6}

We have developed a scenario for fast magnetic reconnection  process for general relativistic magnetohydrodynamical plasmas around Kerr black holes. Generally speaking, the current sheet may form in a non-zero-angular-momentum (non-ZAMO) frame in the black hole background. As a first step toward this, we study magnetic reconnection in the ZAMO frame at first, which corotates with the black hole. We compute the reconnection rate and analyze various important properties of the reconnection layer for two configurations in the equatorial plane of the black hole. We show that compared to the flat spacetime limit, both mass and rotation of the black hole can decrease the reconnection rate. The former (later) dominates for the radial (azimuthal) configuration.

When the current sheet forms in a non-ZAMO frame, we consider a typical example: the Einstein's rotating disk, which orbits around the ZAMO frame with a constant angular velocity. In this case, we studied both the Sweet-Parker and the Petscheck scenario. We established that for any given slow rotations (subrelativistic at most) and mildly relativistic inflow, the ZAMO observer will always find asymmetric reconnection rates for radial configuration: it is decreased on one side of the current sheet and is increased on the other side in comparison to the unrotation limit. This clarifies the effects of rotation on the reconnection layer in the laboratory frame in the flat spacetime limit. These results are valid to a generally rotating reconnection layer, no matter it rotates faster (clockwise) or slower (anti-clockwise) than the ZAMO frame (in the laboratory frame). 

In this work, we focused on the reconnection layer moving in a circular stable orbit in the equatorial plane of the black hole. It will be very interesting to relax the condition and study more general situations, which are relevant to astrophysical events. Our results also have potential applications to energy extraction from black holes via magnetic reconnection, see recent developments in this direction \cite{Comisso:2020ykg}. It is also interesting to consider collisionless effects as well as the gravitational electromotive force in magnetic reconnection. We leave these to future research.

\section*{Acknowledgments}
The work is partly supported by NSFC Grant No. 12275004 and 12205013. Z.Y. Fan was supported in part by the National Natural Science Foundations of China with Grant No. 11805041 and No. 11873025 and also supported in part by Guangzhou Science and Technology Project 2023A03J0016.

\appendix
\section{Sweet-Parker in azimuthal direction for Einstein's disk}
\begin{figure}
\centering
\includegraphics[width=500pt]{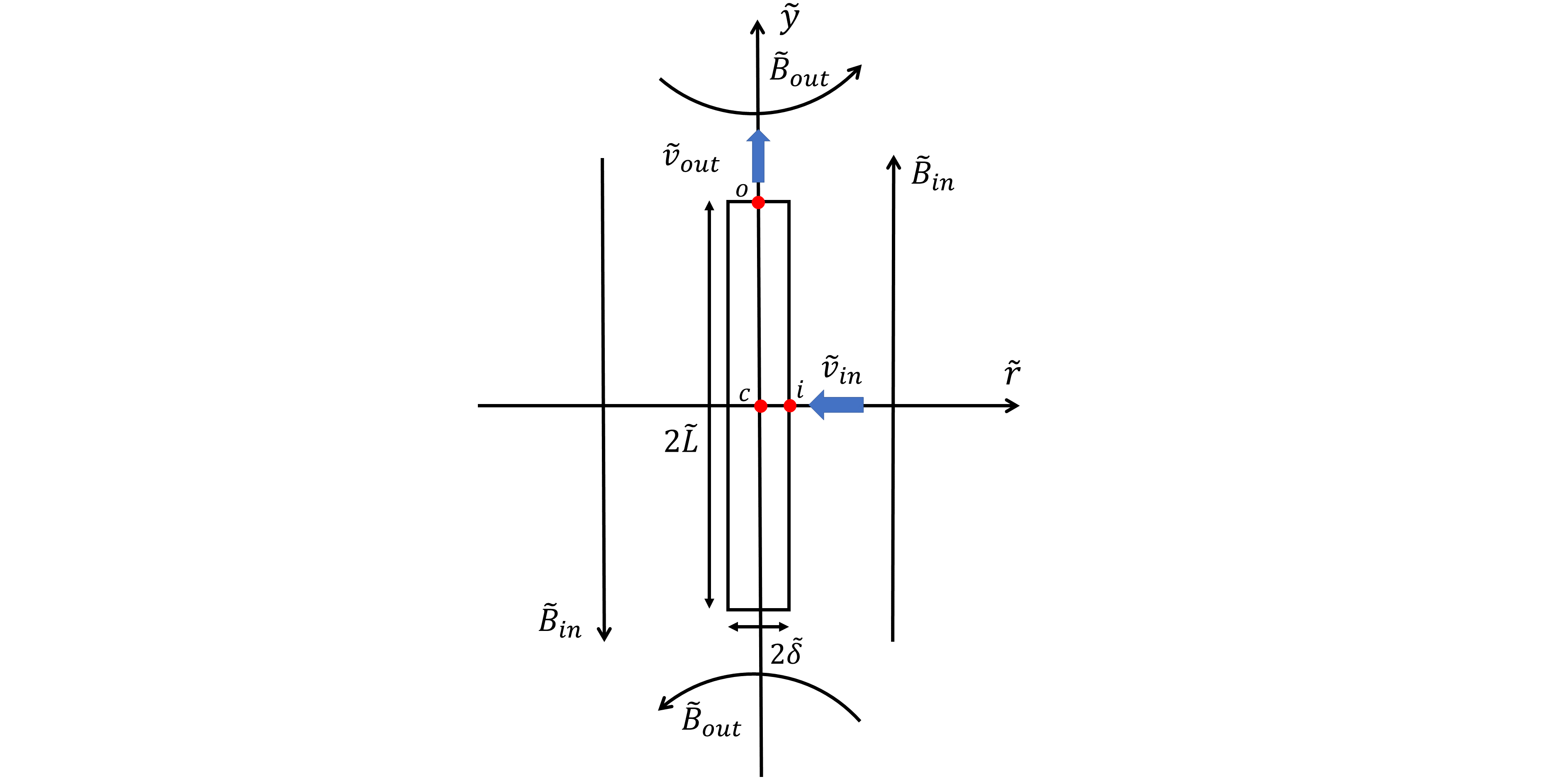}
\caption{Rotating reconnection layer in azimuthal direction in CRF.}
\label{phidisk}
\end{figure}

Consider the Sweet-Parker model in Einstein's disk, with the current sheet located in the $\tilde{y}$-direction. Again we assume $\tilde{v}^z=0=\tilde{B}^z$ and $\tilde\partial_{z}\approx 0$. We set $\tilde{B}_{\mm{in}}=\tilde{B}_0 \tilde e_y$, $\tilde{v}_{\mm{in}}=-\tilde{v}_i\tilde e_r$ and the outflow velocity $\tilde{v}|_O=\tilde{v}_o \tilde e_y$, see Fig. \ref{phidisk}.

From continuity equation, the balance of inflow and outflow flux gives
\be \tilde\delta=\fft{\tilde{\gamma}_i\tilde{v}_i}{\tilde{\gamma}_o \tilde{v}_o}\,\tilde{L}  \,.\label{deltadisk-y}\ee 
Using the Ohm's law, evaluation of the electric field at $i$ and $o$  yields
\bea
&& E_z|_i=q^{-2}\tilde{v}_i\tilde{B}_0 \,,\nn\\
&& E_z|_o=q^{-2}\tilde{v}_o \tilde{B}_o \,.
\eea
Here $\tilde{B}_r|_O\equiv -\tilde{B}_o$ and $q$ is understood to be evaluated at $O$. Since the electric field on the current sheet is uniform, one finds
\be \tilde{B}_o=\fft{\tilde{v}_i \tilde{B}_0}{\tilde{v}_o}\approx  \fft{\tilde{\delta}\tilde{B}_0 }{\tilde{L}} \,,\label{magneticdisk-y}\ee
for mildly relativistic outflow $\tilde{\gamma}_o\approx 1$. Again it is consistent with magnetic flux conservation.

To derive the outflow velocity, we consider the energy-momentum equation in $\tilde{y}$-direction. To leading order, one has
\be\label{disk-emphi} \tilde \partial_{y}\big( h\tilde{\gamma}^2\tilde{v}^2\big)\approx -(q^2\tilde\partial_{ y}p+q\tilde{J}^z \tilde{B}_r) \,.\ee
The current at O can be evaluated as 
\be \tilde{J}^z\Big|_O\approx \tilde\partial_r\tilde{F}^{zr}\Big|_O\approx \fft{\tilde{B}_0}{q\tilde\delta} \,.\ee
Integrating the equation (\ref{disk-emphi}) yields $h\tilde{\gamma}_o^2\tilde{v}_o^2\approx q^2(p+\bar{B}_0^2)$, where $\bar{B}_0=q^{-1}\tilde{B}_0$ is the magnetic field measured by a local observer. Because of $h=4p$ for hot relativistic plasmas and $p=\bar{B}_0^2/2$,
this gives 
\be \tilde{\gamma}_o\tilde{v}_o\approx q \,.\label{outflow-y}\ee
Since $q\leq 1$, the outflow is still mildly relativistic. However, this effects the the reconnection rate and the scaling symmetry significantly. Consider the Ohm's law on the current sheet
\be \tilde{\gamma}(q\tilde{E}_z-q^{-1}\tilde{v}\tilde{B}_r  )=\eta\tilde{J}^z \,.\ee
On the neutral line $\tilde{v}=0$ so that $ \tilde{E}_z|_C\approx \eta \tilde{J}^z/q $. This leads to 
\be \tilde{J}^z\Big|_C\approx \fft{\tilde{v}_i \tilde{B}_0}{q\eta } \,.\ee
Using $\tilde{J}^z|_C\approx \tilde{J}^z|_O$, we obtain 
\be \tilde{v}_i=\fft{\eta}{\tilde{\delta}} \,.\label{vidiskphi}\ee 
Notably, this together with (\ref{deltadisk-y}) and (\ref{magneticdisk-y}) is invariant under the scaling of LRF, which implies $\hat{L}\rightarrow q^{-1}\tilde L\,,\hat{B}_0\rightarrow q^{-1}\tilde{B}_0\,,\hat{v}_o\rightarrow q^{-1}\tilde{v}_o$ and $(\hat\delta\,,\hat{B}_o\,,\hat{v}_i)\rightarrow (\tilde{\delta}\,,\tilde{B}_o\,,\tilde{v}_i)$. 
Combining (\ref{vidiskphi}) with (\ref{deltadisk-y}), we deduce $\tilde\delta=q^{-1/2}\sqrt{\eta\tilde L}$. The reconnection rate for the comoving observer turns out to be
\be
\tilde{v}_i=\Big( \fft{r}{h_3}\Big|_O\Big)^{1/2}\,\,S^{-1/2}\,.
\label{ratedisk-y}\ee
According to the transformation (\ref{bridge}), this is exactly equal to the rate measured by the ZAMO observer, namely $\hat{v}_i=\tilde{v}_i$. Hence unlike the radial case, the rate $\hat{v}_i$ is not changed by rotation of the disk.

\end{document}